\documentclass[a4paper,fleqn,usenatbib]{mnras}

\usepackage{mathptmx}
\usepackage[T1]{fontenc}
\usepackage{ae,aecompl}

\usepackage{graphicx}	
\usepackage{amsmath}	
\usepackage{amssymb}	
\usepackage{multirow}
\usepackage{longtable}
\usepackage{url}
\usepackage{breakurl}
\usepackage{caption}
\usepackage{subfig}
\usepackage{threeparttable}
\usepackage{chngcntr}

\title[4U 1630-472: 2016 and 2018 outbursts]{\textit{AstroSat} and \textit{MAXI} view of the Black Hole binary 4U 1630-472 during 2016 and 2018 Outbursts}

\author[Blessy E. B. et al.]{Blessy E. Baby$^{1,\;2}$\thanks{E-mail: blessy.elizabeth65@gmail.com},
V. K. Agrawal$^{1}$,
Ramadevi M. C.$^{1}$,
Tilak Katoch$^{3}$,
H. M. Antia$^{3}$, 
\newauthor
Samir Mandal$^{4}$, 
Anuj Nandi$^{1}$
\\
$^{1}$Space Astronomy Group, ISITE campus, U. R. Rao Satellite Centre, Karthik Nagar, Bangalore 560037, Karnataka, India\\
$^{2}$Department of Physics, Calicut University, Malappuram 673635, Kerala, India\\
$^{3}$Tata Institute of Fundamental Research, Homi Bhabha Road, Mumbai 400005, Maharashtra, India\\
$^{4}$Department of Earth and Space Sciences, Indian Institute of Space Science and Technology, Thiruvananthapuram 695547, Kerala, India\\
}

\date{Accepted XXX. Received YYY; in original form ZZZ}

\pubyear{2020}

\begin{document}
\label{firstpage}
\pagerange{\pageref{firstpage}--\pageref{lastpage}}
\maketitle

\begin{abstract}
We present an in-depth spectral and timing analysis of the Black Hole binary 4U 1630-472 during 2016 and 2018 outbursts as observed by \textit{AstroSat} and \textit{MAXI}. The extensive coverage of the outbursts with \textit{MAXI} is used to obtain the Hardness Intensity Diagram (HID). The source follows a `c'-shaped profile in agreement with earlier findings. Based on the HIDs of previous outbursts, we attempt to track the evolution of the source during a `super'-outburst and `mini'-outbursts. We model the broadband energy spectra ($0.7-20.0$ keV) of \textit{AstroSat} observations of both outbursts using phenomenological and physical models. No Keplerian disc signature is observed at the beginning of 2016 outburst. However, the disc appears within a few hours after which it remains prominent with temperature ($T_{in}$) $\sim$ 1.3 keV and increase in photon index ($\Gamma$) from 1.8 to 2.0, whereas the source was at a disc dominant state throughout the \textit{AstroSat} campaign of 2018 outburst. Based on the HIDs and spectral properties, we classify the outbursts into three different states - the `canonical' hard and soft states along with an intermediate state. Evolution of rms along different states is seen although no Quasi-periodic Oscillations (QPOs) are detected. We fit the observed spectra using a dynamical accretion model and estimate the accretion parameters. Mass of the black hole is estimated using inner disc radius, bolometric luminosity and two component flow model to be $3-9$ $M_{\odot}$. Finally, we discuss the possible implications of our findings. 
\end{abstract}

\begin{keywords}
accretion -- black hole physics -- stars:individual:4U 1630-47 
\end{keywords}



\section{Introduction}

An in-depth understanding of accretion process in the extreme physical realm of strong gravity can be obtained by studying the X-ray binaries (XRBs), which readily accord us with an astrophysical laboratory. In these systems, a compact object, like a neutron star or a black hole, accretes matter from a companion star. This accretion can take place through stellar winds or Roche lobe overflow in High mass X-ray binaries (HMXB) and Low mass X-ray binaries (LMXB) respectively \citep{Bildsten1997,Corral2016}. Most of the LMXBs are transient in nature wherein the source flux increases above the detection level after a prolonged period of inactivity (i.e., quiescence phase) \citep{Lewin1995,Bildsten1997}. These outbursts can last from weeks to months before the source gradually fades into quiescence. However, the frequency and duration of these outbursts varies from source to source, or sometimes even between different outbursts of the same source itself \citep[and references therein]{Homan2001,Belloni2005,Klis2006,Nandi2012,Sreehari2018}. Detailed spectro-temporal analysis of the XRBs can help us to understand the physical processes which drive these outbursts, providing pointers on the nature of the source itself.

The energy spectrum of a black hole (BH) source is generally modelled by a non-thermal component attributed to a geometrically thick, optically thin corona \citep{Tanaka1995,Narayan1995,Chakrabarti1995} and a thermal component corresponding to a geometrically thin, optically thick Keplerian disc \citep{Shakura1973}. The disc radiates at different temperatures at different radii giving rise to a multi-colour blackbody spectrum, which is considered to be the source of the thermal emission at lower energies. The variation of X-ray flux is tracked in the Hardness Intensity Diagram (HID), which is a plot of intensity with the hardness ratio i.e., the ratio of flux in the hard band to soft band, plotted in the log-log scale \citep{Homan2001}. The HID traces out a `q' shaped profile for a complete outburst in most of the sources \citep[and references therein]{Maccarone2003,Fender2004,Homan2005,Nandi2012,Motta2012,Nandi2018,Radhika2018,Sreehari2019}. Based on the spectral and temporal properties, the sources are seen to evolve from the Low-Hard State (LHS) to the High-Soft Sate (HSS) through the Hard-Intermediate State (HIMS) and the Soft-Intermediate State (SIMS), after which it goes back to the LHS following the reverse order, although the SIMS is not seen during the decay phase for a few sources \citep{Homan2005,Motta2009,Fender2009}. Aperiodic variabilities in the X-ray light curve give rise to narrow features in the Power Density Spectra (PDS) called Quasi-periodic Oscillations (QPOs). C-type QPOs \citep{Casella2005} are almost exclusively found in the LHS and HIMS, whereas A and B-type QPOs are seen mostly in SIMS \citep{Homan2001,Homan2005,Radhika2016} while an absence of QPOs is seen in the HSS.  An alternative state classification proposed by \cite{Remillard2006} only identifies three `stable' states, namely the thermal dominant state (i.e., HSS), LHS and a steep power law (SPL) state. The intermediate states are considered to represent transitions between the `canonical' states. Additional regimes were earlier proposed, before the introduction of the SIMS and HIMS, to explain the case where the inner disc radius does not remain constant as seen in the canonical soft state and the relation between luminosity and disc temperature does not follow the standard $L \propto T_{in}^4$ relation \citep{Kubota2001,Kubota2004,Abe2005}. However, further studies on the properties of the source during these `regimes' in light of recent classification schemes \citep{Remillard2006,Belloni2010}, is hindered by lack of continuous data during outbursts and statistical limitations. 
 
4U 1630-472 is a recurrent X-ray transient discovered by \textit{Uhuru} \citep{Jones1976} while the first detection of an outburst was recorded in 1969 by \textit{Vela 5B} \citep{Priedhorsky1986}. Although the binary orbital period, the nature and mass of its companion remain unkown, its X-ray spectral and timing behaviour indicate that the source is a black hole candidate \citep{Remillard2006}, supporting the earlier classification by \cite{Tanaka1995}. Ever since its discovery, the source has been found to undergo quasi -  periodic outbursts with a recurrence time of approximately 600 days \citep{Parmar1995}. 4U 1630-472 has a high absorption column  density ($N_{H}\sim8\times{10^{22}}$ at. cm$^{-2}$) \citep{Smith2002,Ueda2010} and no optical counterpart has been identified till now, which makes it difficult to estimate the mass and distance of the source. Detection of dips in the 1996 lightcurve suggested a high inclination system with $i = 60-70^{\circ}$ \citep{Kuulkers1998,Tomsick1998,Seifina2014,King2014}. \cite{Augusteijn2001} reported a possible detection of an infrared counterpart during the 1998 outburst, concluding that it is a black hole binary which could be similar to GRO J1655-40 or 4U 1543-47, with a relatively early type star or a Be star as the secondary star. Based on the quiescent luminosity obtained at the end of the 2010 outburst and IR variability, an orbital period greater than $\sim$ 20 hrs ranging upto a few days is expected \citep{Augusteijn2001,Tomsick2014}. An indirect estimation of mass by \cite{Seifina2014} puts the mass of the compact object at $10 \pm 0.1\; M_{\sun}$ based on the relation between the spectral index and mass accretion rate and spectral index and low frequency QPO \citep{Shaposhnikov2009}.  The high absorption places the lower limit of the distance at $>$ 10 kpc while the infrared data analysis reveals the source to lie in the direction of a giant molecular cloud which is at a distance of 11 kpc. By analysing the dust scattering halo of 4U 1630-47 using the 2016 data from \textit{Chandra} and \textit{Swift}, \cite{Kalemci2018} have estimated the distance as 11.5 $\pm$ 0.3 kpc at the far distance estimate and 4.7 $\pm$ 0.3 kpc with the near distance estimate which is less favourable. Reflection continuum modeling suggests that the source hosts a maximally rotating black hole \citep{King2014}. Recently, \cite{Pahari2018} obtained the spin of the compact object as $0.92^{+0.02}_{-0.01}$ by continuum modelling of the spectra obtained using \textit{Chandra} and \textit{AstroSat}, when the source was in the soft state during the 2016 outburst.
\begin{figure*}
    \centering
	\includegraphics[scale=0.35]{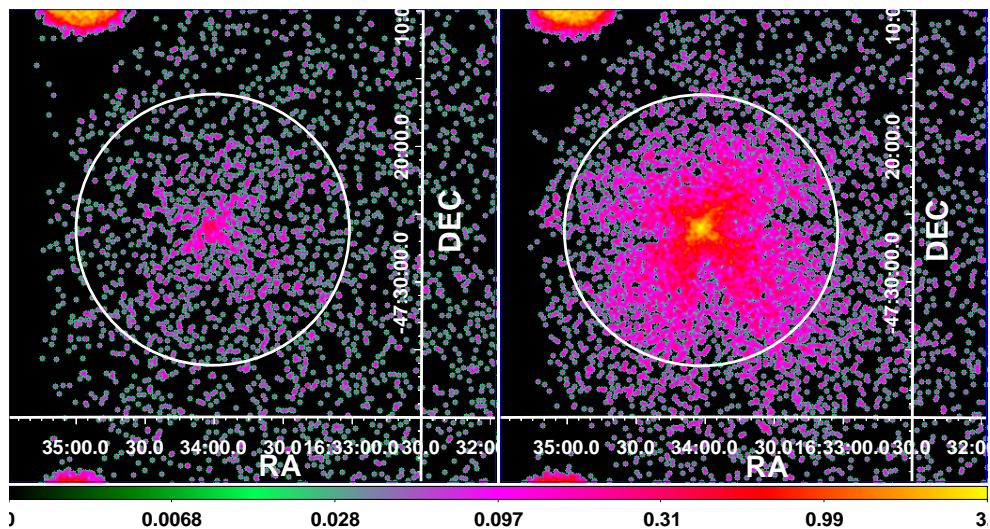}
    \caption{Source region within radius of 10 arcmin is considered for extraction of spectrum for \textit{SXT}. The left image corresponds to Obs 1, where the source is very faint at the beginning of the 2016 outburst. The right image corresponds to Obs 3, which is taken $\sim$ 11 hours apart. It is clear that the source has considerably brightened in this duration. As no pile-up effects were observed, the exclusion of central 1 arcmin region was not performed (see text for details). The colour map at the bottom shows the source intensity for \textit{SXT} in cts pixel$^{-1}$.}
     \label{fig:sxt_src}
\end{figure*}
Till date, 4U 1630-472 has undergone more than 25 outbursts in the $\sim$ 50 years since its detection. At the time of writing this manuscript, the source is found to be in another outburst. Some of the outbursts consist of more than one flaring episode which can last for several years going upto intensities of $\sim$ 0.8 Crab \citep{Tomsick2005} called `super'-outbursts. Four such outbursts have been seen till date \citep{Kaluzienski1978,Parmar1997,Tomsick2005}. The more frequent type of outbursts are the `mini'-outbursts which typically last for $\sim$ $100-200$ days and emit at a peak flux of $< 0.5$ Crab \citep{Abe2005,Tomsick2005}. Lack of bright hard states along with an absence of QPOs is also reported for this source during these outbursts \citep{Abe2005,Tomsick2014,Capitanio2015}. However, the 1998 outburst is an exception as state transitions were clearly seen along with the detection of QPOs \citep{Trudolyubov2001,Seifina2014}, although the outburst lasted only for $\sim$ 100 days. The HID of these outbursts do not follow the typical `q' profile. The 1998 outburst and the `super'-outburst of $2002-2004$ trace out a `c'-shaped profile in the HID \citep{Tomsick2005}. Similar profile is seen for the `mini'-outbursts of 2006, 2008 and 2010 outbursts with quick hard to soft transitions \citep{Capitanio2015}.

 In this work, we undertake a comprehensive analysis of 2016 and 2018 outbursts observed with \textit{AstroSat} and \textit{MAXI}. This provides us the opportunity to perform broadband spectral modelling ($0.7 -20.0$ keV) for this source. We have used the HIDs obtained from \textit{MAXI} and \textit{RXTE/PCA} from the 1998, 1999 and 2002-2004 outbursts to comment on the markers for state transition and attempt to present a global picture for the evolution of the source during the outburst. Based on the HID and the spectral and timing properties, we classify each outburst into different states. The HIDs from previous outbursts are also used to highlight the difference in evolution of the source during a `super'-outburst and `mini'-outburst. We attempt to constrain the mass of the compact object in 4U 1630-472 using the flux obtained during the state transitions. Mass is obtained as a function of a distance where we consider the source to be emitting at Eddington luminosity. Broadband spectra are modelled using two-component advective flow model \citep{Chakrabarti1995,Chakrabarti2006} to understand the accretion flow dynamics. We also attempt to probe the evolution of the source during the `super'-outbursts and `mini'-outbursts based on the HIDs.
 
In Section ~\ref{sec:obs}, we provide the details of observations and the data reduction methods used. In Section ~\ref{sec:anal}, we describe the spectral and timing analysis with the models applied. Results of this analysis along with the outburst profile and HID of both outbursts are presented in Section ~\ref{sec:res} along with the estimation of mass of the black hole by different methods. In Section ~\ref{sec:disc}, we present a comparison of HIDs of different outbursts highlighting their similarities and also comment on the spectral and temporal evolution of the source during a `mini'-outburst. Finally, we summarize our conclusions in Section \ref{sec:conc}. 
\section{Observations and Data Reduction}
 \label{sec:obs}
\textit{AstroSat} \citep{Agrawal2006,Singh2014} observed the source 4U 1630-472 between 2016 August 27 $-$ 2016 October 2 for a total of 4 days. Guaranteed Time (GT) observations were undertaken on August 27 and September 28. Follow up observations on October 1 \& 2 were performed as part of its Target of Opportunity (ToO) campaign.  The subsequent outburst of 4U 1630-47 was observed from 2018 August 4 $-$ 2018 September 17 as part of \textit{AstroSat's} ToO campaign for a total of nine days. The source was monitored continuously by \textit{MAXI} during both outbursts. Continuous observations of the source were also performed by  \textit{Swift/BAT}, except for a few days between 2018 September 9 and September 20. \textit{MAXI} and \textit{Swift/BAT} lightcurves of the source were also extracted for both outbursts.
\subsection{\textit{MAXI} and \textit{Swift/BAT}}

\textit{MAXI} one-day averaged light curve\footnote{\url{http://maxi.riken.jp/mxondem}} in the $2-20$ keV range was analysed to obtain the outburst profile during 2016 and 2018 outbursts. Light curves were also obtained in the ranges $2 - 4$, $4 - 10$ keV, $10 - 20$ keV from MJD 57572 to 58450 to plot the hardness ratios with time. Light curves in the energy range $2-6$ keV and $6-20$ keV were used to obtain the hardness ratio to plot the HID. One-day averaged light curve with Swift/BAT\footnote{\url{https://swift.gsfc.nasa.gov/results/transients/}} in $15 - 50$ keV was extracted to obtain the light curve during both outbursts in the hard energy range.
\subsection{\textit{AstroSat}}
\textit{AstroSat} is a multi-wavelength space observatory equipped with four co-aligned instruments to carry out broadband temporal and spectral studies of cosmic sources, namely, Soft X-ray Telescope (\textit{SXT}) \citep{Singh2014,Singh2017}, Large Area X-ray Proportional Counter (\textit{LAXPC}) \citep{Yadav2016a,Antia2017}, Cadmium Zinc Telluride Imager (\textit{CZTI}) \citep{Vadawale2016} and the Ultra Violet Imaging  Telescope (\textit{UVIT}) \citep{Tandon2017}. Apart from these, the Scanning Sky Monitor (\textit{SSM}) \citep{Seetha2006,Ramadevi2017} scans the sky for X-ray events at all times. In this work, we extensively used broadband observational data from \textit{SXT} and \textit{LAXPC} instruments.

\subsubsection{\textit{SXT}} 
\textit{SXT} has capabilities of X-ray imaging and spectroscopy in the $0.3-8.0$ keV energy range in two modes : Fast Window (FW) mode and the Photon Counting (PC) mode. PC mode data was used for all the observations except for 2016 October $1-2$ where FW mode data was used to avoid pile-up issue in anticipation of higher count rate in the soft energy band as the source was expected to be in the soft state. The Level-1 imaging data is processed off-line through a pipeline software to obtain Level-2 cleaned event files for each individual orbit. 
 \texttt{XSELECT V2.4d} was used in \texttt{HEASOFT 6.21} to extract light curve and spectra using the cleaned event files. We first considered a 16 arc min source region as suggested for PC mode data\footnote{\url{www.tifr.res.in/~astrosat_sxt/dataanalysis.html}}. As the count rate was $<$ 40 cts/s, pile up effects were not observed and hence exclusion of 1 arcmin central source region was unnecessary.  We merged the data from a few orbits in which the count rate and hardness ratios did not vary appreciably to generate the event file using the \texttt{sxtevtmerger} script. For the first day of observation of 2016 August 27, three data groups were obtained separated from the first group by $\sim$ 6 hrs and $\sim$ 11 hrs respectively. We refer to these three data sets as Obs 1 (MJD 57627.05), Obs 2 (MJD 57627.47) and Obs 3 (MJD 57627.69) (see Table \ref{tab:pheno_fit}). The criteria for data selection was a minimum exposure time of 1 ks in \textit{SXT} where the source count rate is atleast a factor of 2 greater than the blank sky background and for which simultaneous observations with \textit{LAXPC} were available. We obtained a total count rate of $\sim$ 0.3 cts/s considering a 16 arcmin source radius for Obs 1. As the source contribution beyond the 10 arcmin radius was $<$ 30\%, we used a 10 arcmin region centred around the source as the optimum case with count rate of $\sim$ 0.23 cts/s. The count rate increased to 1.6 cts/sec for Obs 3. The \textit{SXT} image for Obs 1 and 3 are shown in Figure \ref{fig:sxt_src} along with the chosen source region. It clearly shows that the source brightened by factor $>$ 10 within this duration. For 2016 October $1-2$ FW mode data, 3 arcmin source region was considered encompassing more than 70\% contribution from the source. The response file provided by the \textit{SXT} team for PC mode and FW modes were used\footnote{\url{http://astrosat-ssc.iucaa.in}}, wherever applicable. Appropriate \textit{arf} files provided by the \textit{SXT} team were used where the source was at the centre of the Field of View (FOV). In cases where an offset was observed, \texttt{sxtmkarf} script was used to generate the ancillary response file. Blank Sky observations were used to generate the background spectrum. \textit{SXT} spectra were binned with a grouping of 20 counts per bin for better statistics.   

\subsubsection{\textit{LAXPC}}
\begin{figure*}
	\includegraphics[scale=0.45]{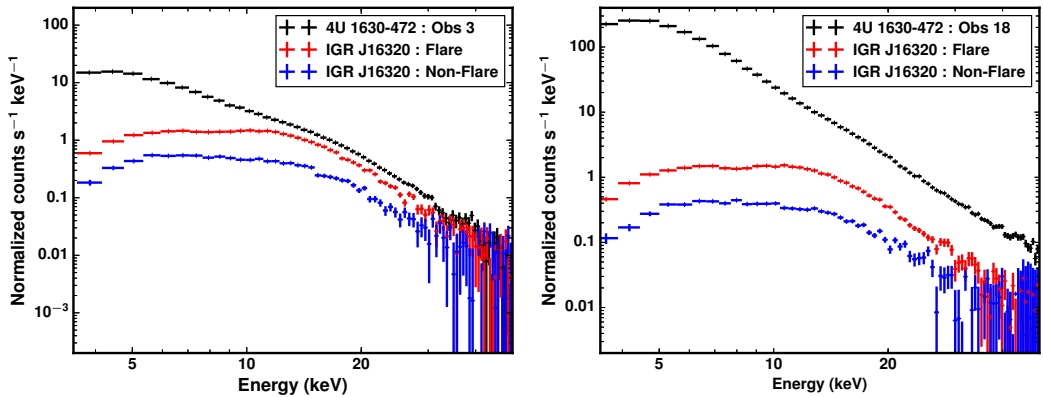}
	\caption{Simulated \textit{LAXPC} spectra for IGR source at an offset of 0.58$^{\circ}$ from the center using the off-axis response plotted for flaring and non-flaring cases \citep{Rodriguez2006} along with the source spectrum obtained for 2016 August 27 (Obs 3) and 2018 September 17 (Obs 18).The model considered for obtaining the simulated spectrum is \textit{wabs*edge*(highecut*powerlaw)}. See text and Appendix \ref{sec:append} for details.}
    \label{fig:off_axis}
\end{figure*}

\textit{LAXPC} consists of three identical X-ray proportional counters (\textit{LAXPC10, LAXPC20} and  \textit{LAXPC30}) with absolute time resolution of 10 $\mu$s in the energy range $3.0 - 80.0$ keV \citep{Yadav2016a,Agrawal2017,Antia2017}. \textit{LAXPC} data is obtained in Event Analysis mode which is analysed using \textit{LAXPC}  software\footnote{\url{http://www.tifr.res.in/~astrosat_laxpc/LaxpcSoft.html}}. We used only \textit{LAXPC20} for the timing and spectral analysis as \textit{LAXPC30} was off and \textit{LAXPC10} was operating at a low gain during the time of the 2018 observations. 
Individual orbits were merged together to satisfy the selection criteria of $>4$ ks for \textit{LAXPC} data, where the count rates and hardness ratios remain stable. From the 2016 and 2018 outburst data, 7 and 11 data sets are respectively chosen based on good exposure time (see Table \ref{tab:pheno_fit}). Due to probable contamination by the nearby source IGR J16320-4751, we do not consider the \textit{LAXPC} data for Obs 1 and 2. Also, depending on the estimated contribution from IGR J16320-4751 source, we restrict the data analysis upto 10 keV, 13 keV or 20 keV for different ObsIDs (see Section \ref{sec:contamination} and Appendix \ref{sec:append} for details). We used \texttt{laxpcsoftv2.5} released on 2018 August 6,  to obtain the lightcurve and energy spectra. \textit{LAXPC} has proved an invaluable resource in aiding the spectro-temporal studies of various sources \citep{Misra2017,Sreehari2019}. We follow these papers and the instructions provided with the \textit{LAXPC} data reduction software. We consider only the top layer of the detector and discard multiple detections of a single event to obtain the spectrum. Details of the response and background generation are found in \cite{Antia2017}. Background was generated with the gain shift applied to align the background spectra with the source spectra using the \texttt{backshiftv2.e} code. Different background models are provided by the \textit{LAXPC} team for each month of operation, from which the background model closest to the observation dates were chosen. However, four different backgrounds were provided for the month of August 2018 due to variation in background rate, corresponding to August 2, 8, 12 and 15. The background for 2018 August 15 was chosen for all the observations. The details of the ObsIDs and exposure times of the observations are presented in Table \ref{tab:pheno_fit}. Analysis and modelling procedure of \textit{LAXPC} spectra is further elaborated in Section \ref{sec:anal}.

\subsection{Contamination of \textit{AstroSat} data by nearby persistent source IGR J16320-4751}
 \label{sec:contamination}

4U 1630-472 lies in a crowded region of Galaxy where hard X-ray sources are found in close proximity to each other \citep{Rodriguez2003}. \textit{LAXPC} on-board \textit{AstroSat} has a collimated FOV of $0.9^{\circ} \times 0.9^{\circ}$ at FWHM \citep{Antia2017}. Thus, it is essential to ensure that the data obtained during \textit{AstroSat} campaign was free from such a contamination. Images obtained with \textit{SXT} were checked for each orbit to confirm that other sources within the \textit{SXT} FOV did not contribute to the low energy flux ($0.3-8$ keV). There are two persistent sources close to 4U 1630-472, namely, IGR J16320-4751 ($=$AX J1631.9-4752) and IGR J16393-4643, which could contaminate the \textit{LAXPC} spectrum. IGR J16393-4643 lies at an angular separation of more than 1$^{\circ}$ from 4U 1630-472 and hence is excluded as a possible candidate for contamination. However, as IGR J16320-4751 (hereafter referred to as IGR) lies 0.58$^{\circ}$ away from 4U 1630-4751 \citep{Rodriguez2006}, this source lies close to the edge of the \textit{LAXPC} FOV at FWHM. The contribution could be more significant in the higher energy band as the FOV is quoted to be $1^{\circ} \times 1^{\circ}$ at $E > 50$ keV \citep{Antia2017}. Therefore, we adopt a two way approach to deal with this issue. The first is using pulsation studies to confirm the detection of the IGR source along with its relative strength (see Appendix \ref{sec:append} for details). The second approach is to quantify the contribution from this source to the total source spectrum by generating simulated spectrum using an off-axis response based on the previous studies of IGR source.

IGR is a hard, persistent source thought to be a pulsar emitting at a pulse period of $\sim$ 1310 s \citep{Rodriguez2006}. Therefore, it is likely that this source contributes at the higher energy spectra obtained using \textit{LAXPC}. Energy dependent lightcurves of both sources were obtained from the \textit{Swift/BAT} team on request as both sources remain quiescent for most of the duration of 2016 and 2018 outbursts (private communication with Amy Lien and Hans Krimm). We also looked for simultaneous data for the IGR  source using different high energy instruments. Although \textit{INTEGRAL} observations exist during August 2016 for both sources, they cannot be used for detailed spectral analysis as the observations are in slew mode (private communication with Carlo Ferrigno).

Therefore, for all the observations, we check for the reported $\sim$ 1300 s pulsation in the $3-80$ keV \textit{LAXPC} lightcurve. This is used to obtain the pulse fraction from which the probable contribution from the IGR source is estimated. The details of the pulsation studies and the amplitude strengths are provided in Appendix \ref{sec:append}. Constraints on pulsation could not be obtained for Obs 1 \& 2 due to less exposure time and low counts. For all the other observations, pulsation studies proved beyond doubt that the spectrum was contaminated by the IGR source at varying levels for different observations. We consider the ObsIDs for which the percentage of contribution from IGR source is found to be less than 5\% to be relatively free from contamination. However, to further consolidate our results, we generate the simulated spectrum for IGR considering the unlikely flaring case with a $2-10$ keV flux of $2.4 \times 10^{-10}$ ergs cm$^{-2}$ s$^{-1}$ \citep{Rodriguez2006} using the off-axis response and compare it with the spectra obtained from \textit{LAXPC}. The model used to generate the simulated spectrum is \textit{wabs*edge*(highecut*powerlaw)}. The expected count rate for the IGR source using off-axis response in the $3-80$ keV band was close to 20 cts/s for the flaring case. This corresponds to a contribution of $>$ 20\% for Obs 1 \& 2 where the count rates are very low. Hence, we choose to discard \textit{LAXPC} data for Obs 1 \& 2. For the rest of the observations, we check for the relative contribution from the IGR source for each observation individually. In Figure \ref{fig:off_axis}, we present two cases. The first corresponds to the case where the source remained quiet in the $15-50$ keV range as seen in Figure \ref{fig:maxi_fig} and second corresponds to the high energy activity seen in \textit{Swift/BAT} lightcurve. The spectra for the non-flaring case is also shown for comparison. However, we consider the unfavourable flaring case in order to exclude any possibility of contamination from the IGR source. It is clearly seen from Figure \ref{fig:off_axis} that the IGR source dominates above 13 keV in the first case (for e.g., Obs 3, left panel) and above 25 keV in the second (for e.g., Obs 18, right panel). Therefore, for the ObsID's where the contamination is less than 5\% (see Table \ref{tab:app1} in Appendix \ref{sec:append}), we consider \textit{LAXPC} spectra upto 13 keV exercising extreme caution. We also perform spectral fitting for the source spectrum considering the simulated IGR spectrum as background. It was found that the spectral parameters remain unchanged upto 13 keV for cases where the IGR contribution is $\leq$ 10\% in comparison with the constant background subtracted spectra and upto 10 keV for cases where IGR contribution is more than 10\%. For the last two ObsIDs, we could obtain good spectra upto 20 keV considering the brightness of the source in the hard energy band and the IGR spectra subtraction method (see Figure \ref{fig:maxi_fig} and Tables \ref{tab:app1}, \ref{tab:app2}). Therefore, for spectral analysis, we consider \textit{LAXPC} spectra upto 10 keV for Obs 3, 10-12 \& 16, upto 13 keV for Obs 4-7, 8-9, 13-15 and upto 20 keV for Obs 17-18 (see Table \ref{tab:app1} for a summary  on source contribution and consideration of energy bands). However, for uniformity, timing analysis is performed in the $3-10$ keV range.

\section{Analysis and Modelling Procedure}
 \label{sec:anal}
\subsection{Spectral Analysis}
\textit{AstroSat} provides us an opportunity to model the broadband spectra of 4U 1630-472 source. To understand the evolution of the spectral parameters with time during 2016 and 2018 outbursts, we attempt to model the spectra using phenomenological and physical models. \texttt{XSPEC} version 12.9.1 is used to perform the fits. 

The combined spectra of \textit{SXT} and \textit{LAXPC} were modelled in three different energy bands for different ObsIDs i.e., $0.7-10$ keV, $0.7-13$ keV and $0.7-20$ keV based on the pulsation studies and estimates of relative contribution from IGR source (see Table \ref{tab:app1} and Table \ref{tab:pheno_fit}). The \emph{gain} command is used to modify the \textit{SXT} response file linearly. The slope is fixed to 1 and the offset is allowed to vary. Obs 1 \& 2 are modelled using \textit{phabs $\times$ powerlaw} model using only \textit{SXT} data as \textit{LAXPC} data are contaminated by the IGR source as mentioned in Section \ref{sec:contamination}. No additional disc component was required implying that the source was in hard state. The $N_{H}$ value was obtained at 7.5 $\times$ 10$^{22}$ atoms cm$^{-2}$ with a photon index of 1.82 and $\chi_{red}^{2}$ of 32.1/27 $=$ 1.2. The photon index increased to 1.91 for Obs 2. A disc component is required for Obs 3 with photon index increasing to 2.03. For the rest of the 2016 observations and all of 2018 observations, we used an absorbed multicolour disc component model (\textit{diskbb}) \citep{Mitsuda1984,Makishima1986}, a power law component (\textit{powerlaw}) and a \textit{gaussian} to fit the spectra. Power law component was not required in the observations where spectrum was considered only upto 10 keV due to prominence of disc. The $N_{H}$ was obtained at ($7.3-8.13$) $\times$10$^{22}$ atoms cm$^{-2}$. An edge was required at 2.44 keV for Sulphur\footnote{\url{http://www.iucaa.in/~astrosat/kps_a01_sxt.pdf}} in addition to the gain applied for \textit{SXT} data.

A systematic error of 1.5\% is applied to each bin in the combined fit. A normalization constant is added to account for the difference in calibration of the instruments. The centroid energy of the \textit{gaussian} is restricted to be in the range $6.2-6.9$ keV, wherever the component was required. The fit values obtained from this model are given in Table~\ref{tab:pheno_fit}. All the errors are quoted at 68\% confidence unless stated otherwise.

\subsection{Timing Analysis}
Lightcurves of 1 s bin were obtained in the $3-10$ keV range using \textit{LAXPC20} initially for all observations considered in Table \ref{tab:pheno_fit}.  For these combined orbits, we generated 0.005 s resolution lightcurves and corresponding backgrounds. The background varies between $\sim$ $6-10$ cts/sec during the observations. This background was subtracted from the lightcurve to generate the final lightcurve. We divide the lightcurve into 16384 time bins and create a power spectra for each of the intervals. Then the power spectra are averaged to generate the final PDS. We rebin the power spectra geometrically by a factor of 1.05 in the frequency space. Dead time corrected Poisson noise level was obtained following the procedure in \cite{Agrawal2018,Sreehari2019} using the relation and values given in \cite{Zhang1995} and \cite{Yadav2016a}. This dead time corrected Poisson noise subtraction is applied to all the PDS, which are normalized to give fractional rms spectra (in units of (rms/mean)$^{2}$/Hz)). No contribution was seen in the PDS beyond 10 Hz for all the observations. All the PDS were modelled with a power law in the range 0.005 to 10 Hz (see Figure \ref{fig:pds}). Red noise is dominant in the PDS with no significant power above $\sim$ 1 Hz. No QPOs were observed. During the 2016 observations, the source count was too low in the initial observations resulting in noisy PDS making it difficult to check for the presence of QPOs. Total rms values were obtained in the frequency range $0.005-10$ Hz. Although the \textit{LAXPC} data for Obs 2 is discarded due to possible contamination by the nearby IGR source (see Section \ref{sec:contamination}), the rms value for this particular observation is presented for continuity sake. However, the number quoted is only for reference and should be treated thus. The outburst profile and HID of the outbursts along with spectral and temporal properties of the source are presented in the next section.
\section{Results}
 \label{sec:res}
\subsection{Outburst Profile and HID}
  \label{sec:hid}
\begin{figure*}
	\includegraphics[scale=0.55]{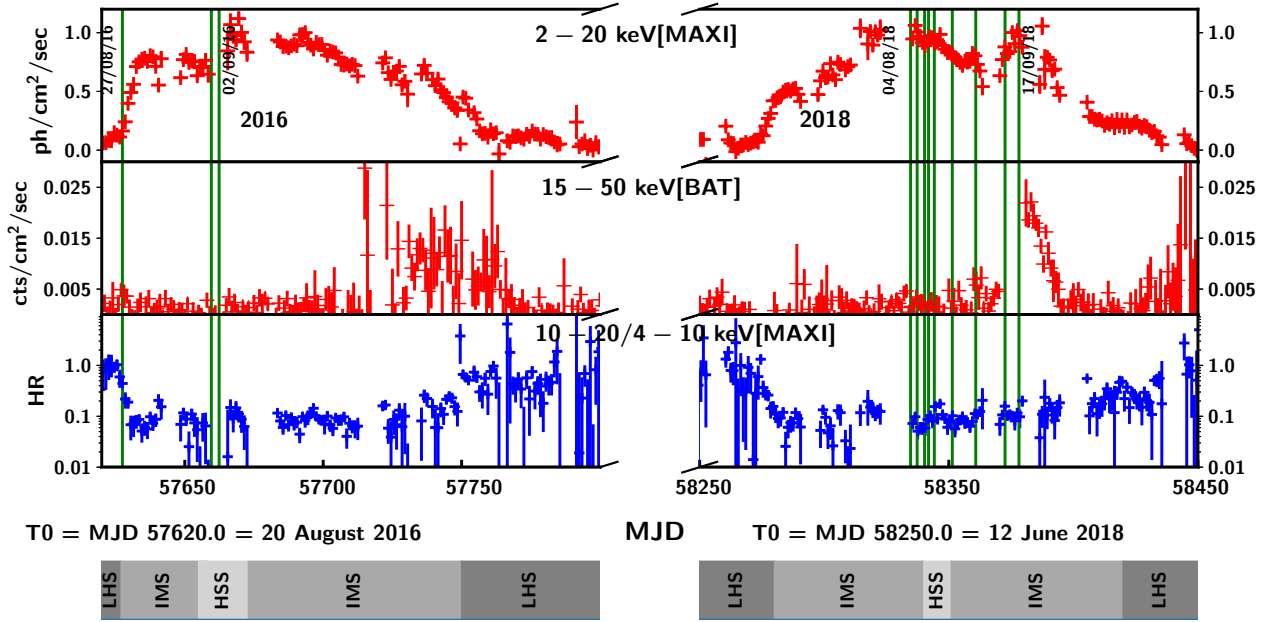}
    \caption{Top panel shows the \textit{MAXI} $2-20$ keV lightcurve of the outburst. The second panel shows the \textit{Swift/BAT} lightcurve in $15-50$ keV range. Hardness ratio obtained from \textit{MAXI} is plotted in panel 3. \textit{AstroSat} observations available for both outbursts of 2016 and 2018 are marked as vertical green lines. The colour bar at the bottom indicates different states. See text for details.}
     \label{fig:maxi_fig}
\end{figure*}
\begin{figure}
	\includegraphics[scale=0.6]{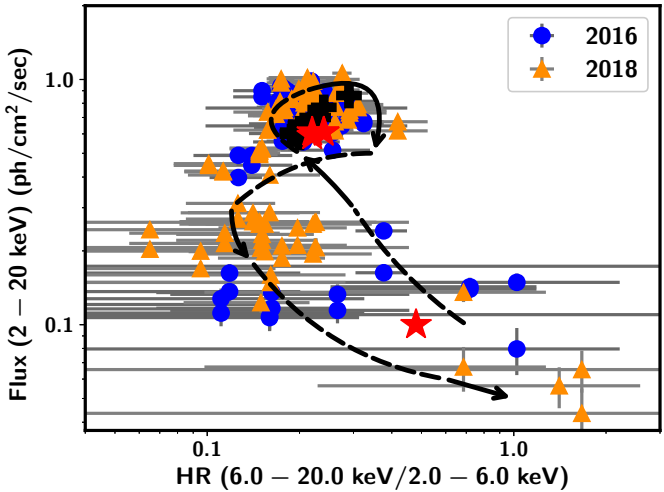}
    \caption{HID obtained with \textit{MAXI} observation of 2016 and 2018 outbursts. The blue circles and orange triangles correspond to \textit{MAXI} data of 2016 and 2018 outbursts respectively.The uncertainties are represented in grey. Hardness ratio is taken as ($6.0-20.0$ keV/$2.0-6.0$ keV). The red stars and black filled crosses represent the data obtained using \textit{AstroSat} during the 2016 and 2018 outbursts respectively. The red star at the bottom corresponds to Obs 3 from 2016 outburst. The arrows represent the direction of evolution throughout the outbursts.Note that the HR defined here is different than the one used in Figure \ref{fig:maxi_fig} (see text for details).}
     \label{fig:maxi_hid}
\end{figure}
To get a comprehensive picture of the `mini'-outbursts of 4U 1630-472, we analyse its lightcurve in different energy bands using \textit{MAXI} and \textit{Swift/BAT}. The \textit{MAXI} lightcurve ($2 - 20$ keV) of the 2016 and 2018 outbursts of the source is shown in the top panel of Figure \ref{fig:maxi_fig}. The 2016 outburst was first detected on 2016 August 27  and the source was active for $\sim$ 100 days before going into quiescence. After $\sim$ 600 days of inactivity, the source again went into outburst on 2018 June 4 which lasted for $\sim$ 150 days. Both 2016 and 2018 outbursts show similar profiles. The days of \textit{AstroSat} observations are marked as vertical green lines in Figure \ref{fig:maxi_fig}.

The flux variation observed with \textit{Swift/BAT} during both outbursts is shown in Panel 2 of Figure \ref{fig:maxi_fig}. The source was very faint at the beginning of the outburst, but the indication of a sudden rise in \textit{Swift/BAT} counts is seen in the decay phase of the outburst. This increase is more pronounced in the 2018 outburst than the 2016 outburst, which also precedes the secondary peak observed during the 2018 outburst. The flux increased $\sim$4 times in one day from MJD 58360 to MJD 58361. Observations are not available for the 2018 outburst between MJD 58370 and MJD 58381 due to some technical snag in \textit{Swift/BAT} (private communication with Hans Krimm). Panel 3 of Figure \ref{fig:maxi_fig} shows the hardness ratio variation with time. The hardness ratio (HR) is defined as the ratio of flux in $10-20$ keV band to the flux in $4-10$ keV band. At the beginning of the 2016 outburst, the HR values are closer to $\sim$ 1.0 and reduce to $\sim$ 0.2, showing a quick transition from hard to an intermediate state. Then the source goes into the HSS as shown in the color bar at the bottom of Figure \ref{fig:maxi_fig}. The source then gradually decays to the LHS with a increase in HR values, through a relatively longer intermediate state. Similar trend is followed in the 2018 outburst, with a longer rise time than in the 2016 outburst. The source seems to evolve from hard to soft state through an intermediate state during rise. The source stays in the HSS for a few days. There is a rise in hardness ratios from $\sim$ 0.06 to $\sim$ 0.1 accompanied by the onset of secondary outburst, where the source is in the intermediate state. Finally, the source goes into the LHS with a decrease in flux and increase in hardness ratios. The HID of both outbursts is studied further to confirm the state transitions as inferred from Figure \ref{fig:maxi_fig}.
\begin{figure*}
	\subfloat{%
	\hspace{0cm}
	\includegraphics[scale=0.45]{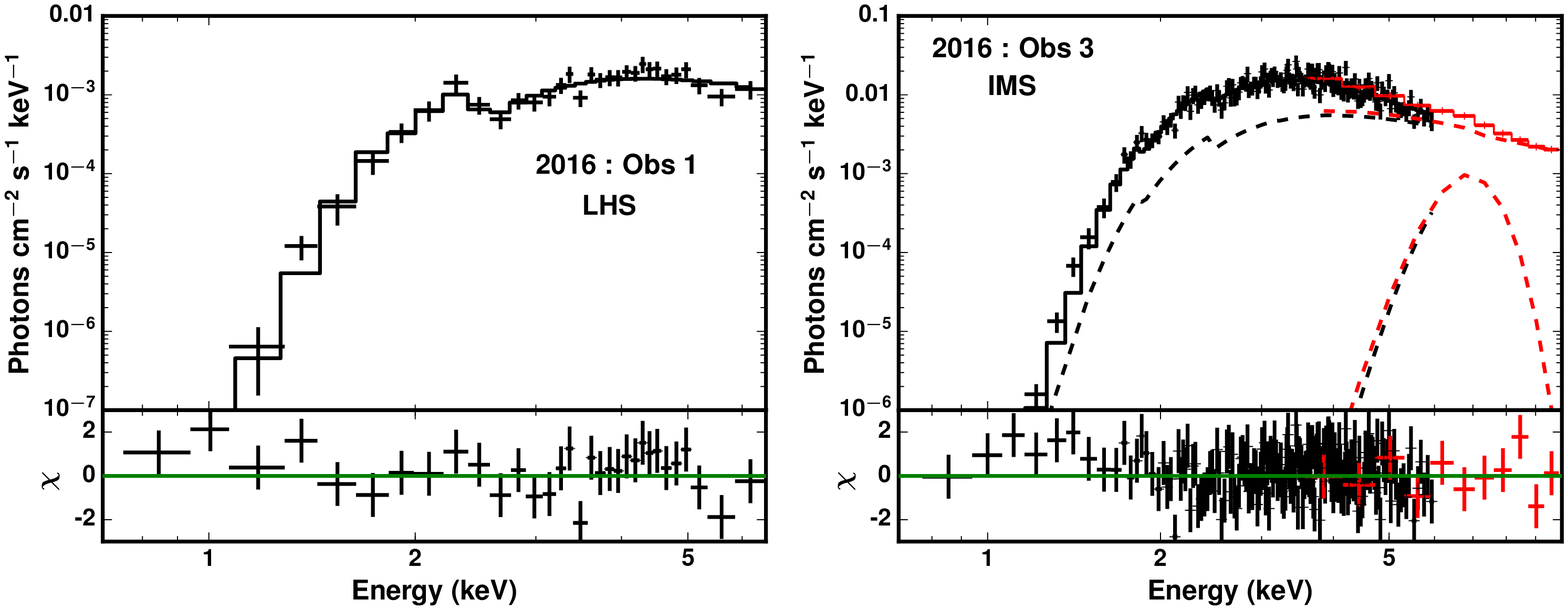}}
    \hspace{0cm}
    \subfloat{%
	\includegraphics[scale=0.45]{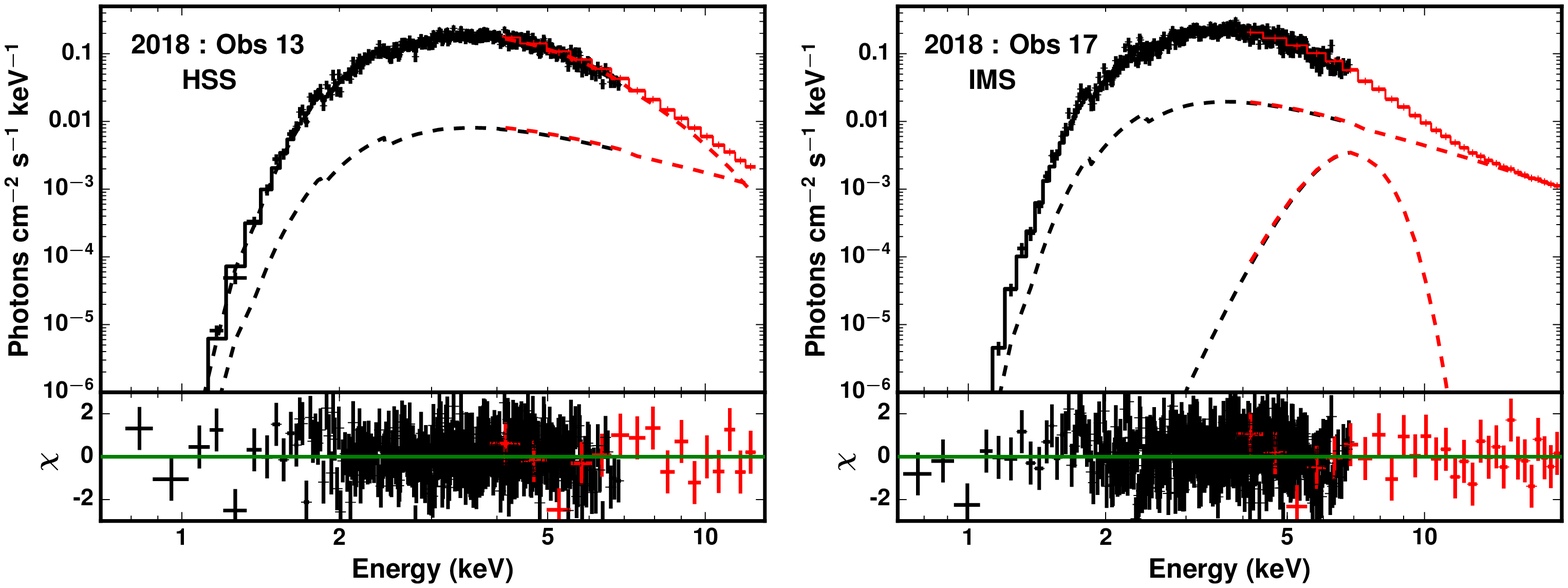}}
    \caption{Broadband spectra from \textit{SXT} and \textit{LAXPC} are shown for two sample observations each from 2016 and 2018 outbursts for LHS, IMS, HSS and IMS. \textit{SXT} data is shown in black and \textit{LAXPC} data in red. For Obs 1 (2016), only \textit{SXT} data has been modelled. }
    \label{fig:pheno_fit_fig}
\end{figure*}   
In Figure \ref{fig:maxi_hid}, HID obtained with \textit{MAXI} for 2016 and 2018 outbursts is shown. Note that a different definition of hardness ratio is used to generated the HID from the one used in Figure \ref{fig:maxi_fig}. While the third panel in Figure \ref{fig:maxi_fig} was used as a marker for state transition, the HID is solely used to track the evolution of the source during the outburst. Due to larger uncertainties in the flux in $2-4$ keV band in \textit{MAXI} data for 2016, negative flux is reported in some of the cases resulting in loss of data points during the initial phase of 2016 outburst. Therefore the harder energy bands are used for better representation in Panel 3 of Figure \ref{fig:maxi_fig}. To track the evolution of the source in the HID, we include lower energy bands at the cost of loss of some data points. The observations where negative fluxes are reported in either of the energy bands ($2-6$ keV and $6-20$ keV) are discarded. The blue circles and orange triangles represent the \textit{MAXI} data during the 2016 and 2018 outbursts respectively. The uncertainties are represented in grey. Data points obtained using \textit{AstroSat} are superimposed on it using the red stars and black filled crosses for the 2016 and 2018 outbursts respectively. The hardness ratio decreases from $\sim$ 0.6 to $\sim$ 0.2 during the rise of the 2016 outburst along with an increase in luminosity represented by blue circles. The red star is the first day of \textit{AstroSat} observation of the source on 2016 August 27 corresponding to Obs 3,  when it is in a harder state. As the outburst progresses, there is a quick transition to the HSS accompanied by an increase in luminosity. This quick change in the HR values can also be seen in Panel 3 of Figure \ref{fig:maxi_fig} for the 2016 outburst. There is a hint of further increase in luminosity along with an increase in hardness ratios, suggesting a sudden transition into an intermediate state. The source then reaches the LHS through the path indicated by the arrows ending at a lower flux than that at the beginning of the outburst. Similar trend is seen for the 2018 outburst represented by orange triangles. 
The HIDs do not follow the typical `q'-profile observed for Low mass BH XRBs as is already observed in previous outbursts of this source \citep{Abe2005,Capitanio2015} rather a distorted `c'-shaped profile as traced out in Figure \ref{fig:maxi_hid}. The source exists in the LHS at the bottom of the `c', while the HSS is almost exclusively found at the top end. However, intermediate states (IMS) seem to exist everywhere except the lower end of the HIDs. Therefore, we attempt to understand the spectral and timing evolution of the source during these states in the following sections.
\subsection{Spectral Properties}
 \label{sec:spec_prop}

As discussed in Section \ref{sec:anal}, we fitted the broadband spectra with phenomenological models. Figure \ref{fig:pheno_fit_fig} shows broadband spectra from 2 different observations each from 2016 and 2018 outbursts. The top left spectra shows the fit to Obs 1. No disc component is required, indicating that the source was in a `canonical' hard state which is also inferred from the HR values in Figures \ref{fig:maxi_fig} and \ref{fig:maxi_hid}. For Obs 3, separated by 11 hrs, the disc component is required to obtain a good fit ($\chi_{red}^{2}$ = 1.09) with a disc temperature of 0.79 keV and a normalization of $\sim$ 179 (see Table \ref{tab:pheno_fit} and top right spectra in Figure \ref{fig:pheno_fit_fig}). Photon index ($\Gamma$) increased from 1.8 to 2.0 during the first 3 observations of 2016 observations after which the disc became more prominent. The disc temperature increased from 0.79 keV to 1.36 keV accompanied by rise in photon index, signalling the transition from hard to soft state. For the 2018 outburst, the disc temperature was in the range $1.25-1.38$ keV with power law index reducing to from 2.8 to 2.09 in the last three observations where the source seems to have transitioned to brighter IMS corresponding to the rise in \textit{BAT} lightcurve (see Figure \ref{fig:maxi_fig} and bottom spectra in Figure \ref{fig:pheno_fit_fig}).    

\begin{table*}
	\centering
	\begin{threeparttable}[b]
	\caption{Fit parameters using the model \textit{phabs $\times$ (diskbb+gaussian+powerlaw)}. The errors are quoted with 68$\%$ confidence.}
	  \label{tab:pheno_fit}
	\begin{tabular}{c@{\hspace{2pt}}c@{\hspace{2pt}}cc@{\hspace{2pt}}c@{\hspace{2pt}}cccccc@{\hspace{2pt}}c} 
		\hline
		Date & Obs & MJD & \multicolumn{2}{c|}{Eff. Exp. (ks)} & $N_H$ & $\Gamma$ & $E_{line}$ & $T_{in}$ & $Norm_{disk}$ & $Flux_{0.7-20.0}$ & $\chi^2_{red}$\\
		(ObsID) &  &  & \textit{SXT} & \textit{LAXPC} & (at./cm$^{2}$) &  & (keV) & (keV) & & $(\times 10^{-8} ergs\;cm^{-2}\;s^{-1}$) & ($\chi^{2}/dof$)\\
		\hline
		& & & & & & 2016 & & & & \\
		\hline
		Aug 27 (0626) & 1 & 57627.05 & 3.8 & - & $7_{-1}^{+2}$ & $1.8_{-0.6}^{+0.7}$ & - & - & - & $0.030 \pm 0.003$ & 1.2 (32.1/27) \\ 
		 & 2 & 57627.47 & 5.3 & - & $7.3_{-0.6}^{+0.5}$ & $1.9_{-0.2}^{+0.1}$ & - & - & - & $0.081 \pm 0.001$ & 1.24 (145/116) \\
         & 3 & 57627.69 & 3.8 & 10.4 & $8.5_{-0.2}^{+0.2}$ & $2.0_{-0.2}^{+0.1}$ & $6.8^{l}$ & $0.79_{-0.01}^{+0.03}$ &  $179_{-16}^{+39}$ & $0.27 \pm 0.01$ & 1.07 (192/179) \\ 
        Sep 28 (0686) & 4 & 57659.04 & 5.1 & 9.6 & $7.2_{-0.1}^{+0.1}$ & $2.25_{-0.03}^{+0.02}$ & $6.3^{l}$ & $1.36_{-0.01}^{+0.01}$ & $222_{-9}^{+10}$ & $1.79 \pm 0.01$ & 1.20 (447/372) \\ 
        Sep 29 (0686) & 5 & 57660.14 & 1.0 & 9.1 & $7.5_{-0.2}^{+0.2}$ & $2.65_{-0.02}^{+0.02}$ & $6.8^{l}$ & $1.32_{-0.01}^{+0.01}$ & $206_{-14}^{+16}$ & $1.55 \pm 0.02$ & 1.16 (392/338) \\ 
        Oct 1 (0698) & 6 & 57662.40 & 2.5 & 5.6 & $7.9_{-0.1}^{+0.2}$ & $2.44_{-0.01}^{+0.01}$ & $6.2^{l}$ & $1.35_{-0.01}^{+0.01}$ & $227_{-12}^{+16}$ & $1.54 \pm 0.02$ & 1.32 (519/391) \\
        Oct 2 (0698) & 7 & 57663.03 & 1.9 & 9.8 & $7.8_{-0.2}^{+0.1}$ & $2.38_{-0.03}^{+0.03}$ & $6.2^{l}$ & $1.36_{-0.01}^{+0.01}$ & $235_{-17}^{+16}$ & $1.62 \pm 0.01$ & 1.20 (401/335) \\
        \hline
        & & & & & & 2018 & & & & \\
		\hline
        Aug 4 (2274) & 8 & 58334.55 & 1.9 & 4.2 & $7.9_{-0.2}^{+0.1}$ & $2.19_{-0.01}^{+0.01}$ & - & $1.29_{-0.004}^{+0.01}$ & $296_{-18}^{+6}$ & $1.74 \pm 0.01$ & 1.2 (541/444)\\ 
         & 9 & 58334.75 & 2.6 & 9.4 & $7.9_{-0.1}^{+0.1}$ & $2.25_{-0.01}^{+0.01}$ & - & $1.30_{-0.01}^{+0.003}$ & $291_{-8}^{+7}$ & $1.79 \pm 0.02$ & 1.2 (549/467)\\
        Aug 7 (2282) & 10 & 58337.29 & 1.5 & 6.1 & $7.5_{-0.1}^{+0.1}$ & - & - & $1.38_{-0.01}^{+0.01}$ & $224_{-7}^{+8}$ & $1.58 \pm 0.02$ & 1.2 (509/422) \\
        Aug 10 (2294) & 11 & 58340.27 & 1.3 & 6.5 & $7.8_{-0.2}^{+0.1}$ & - & - & $1.33_{-0.003}^{+0.01}$ & $256_{-12}^{+6}$ & $1.56 \pm 0.02$ & 1.1 (446/391) \\
        Aug 11 (2298) & 12 & 58341.86 & 2.8 & 8.1 & $7.7_{-0.1}^{+0.1}$ & - & - & $1.34_{-0.003}^{+0.0}$ & $244_{-8}^{+4}$ & $1.48 \pm 0.01$ & 1.2 (597/473) \\
        Aug 14 (2304) & 13 & 58344.18 & 1.2 & 9.5 & $7.8_{-0.2}^{+0.1}$ & $2.22_{-0.01}^{+0.02}$ & - & $1.26_{-0.01}^{+0.01}$ & $306_{-18}^{+13}$ & $1.58 \pm 0.02$ & 0.98 (370/376) \\
        Aug 20 (2318) & 14 & 58350.86 & 2.5 & 7.5 & $8.08_{-0.05}^{+0.02}$ & $2.79_{-0.01}^{+0.01}$ & - & $1.26_{-0.01}^{+0.01}$ & $282_{-3}^{+6}$ & $1.49 \pm 0.02$ & 1.2 (567/462) \\
        Aug 21 (2318) & 15 & 58351.94 & 1.3 & 7.2  & $7.7_{-0.1}^{+0.1}$ & $2.11_{-0.01}^{+0.02}$ & - & $1.25_{-0.01}^{+0.01}$ & $300_{-8}^{+8}$ & $1.54 \pm 0.01$ & 1.05 (405/384) \\
        Aug 30 (2340)  & 16 & 58360.81 & 2.4 & 7.3 & $7.5_{-0.1}^{+0.1}$ & - & - & $1.35_{-0.01}^{+0.01}$ & $217_{-9}^{+7}$ & $1.37 \pm 0.02$ & 1.3 (623/451) \\
        Sep 11 (2354) & 17 & 58372.57 & 1.0 & 6.6 & $8.1_{-0.1}^{+0.1}$ & $2.16_{-0.02}^{+0.04}$ & $6.8^{l}$ & $1.26_{-0.01}^{+0.01}$ & $355_{-13}^{+22}$ & $2.11 \pm 0.02$ & 1.04 (448/431) \\ 
        Sep 17 (2372) & 18 & 58378.16 & 1.2 & 6.4 & $7.7_{-0.1}^{+0.1}$ & $2.09_{-0.01}^{+0.01}$ & $6.8^{l}$ & $1.28_{-0.02}^{+0.01}$ & $314_{-13}^{+40}$ & $2.28 \pm 0.03$ &  1.26 (523/414) \\ 
		\hline
	\end{tabular}
	\begin{tablenotes}
	 \item[$l$] Parameter pegged at bounds
	\end{tablenotes}
   \end{threeparttable}
\end{table*}
All the $N_{H}$ values lie between (7$-$8.5)$\times$10$^{22}$ atoms cm$^{-2}$ (see Table \ref{tab:pheno_fit}). The photon index increases from 1.8 to $\sim$ 2.0 as the source progresses from the LHS to IMS, with the disc emerging in prominence thereafter as it transits to HSS during the 2016 outburst. The disc temperature remains fairly constant ($1.25-1.38$ keV) for all the observations except for the IMS in 2016. In Section \ref{sec:disc}, we comment on the evolution of the source during different states and attempt to explain it in the context of other physical models.
\subsection{Temporal Properties}
 \label{sec:temp_prop}
\begin{figure*}
  \begin{minipage}{0.5\textwidth}
   \subfloat[b][\label{fig:pds}]{\includegraphics[scale=0.45]{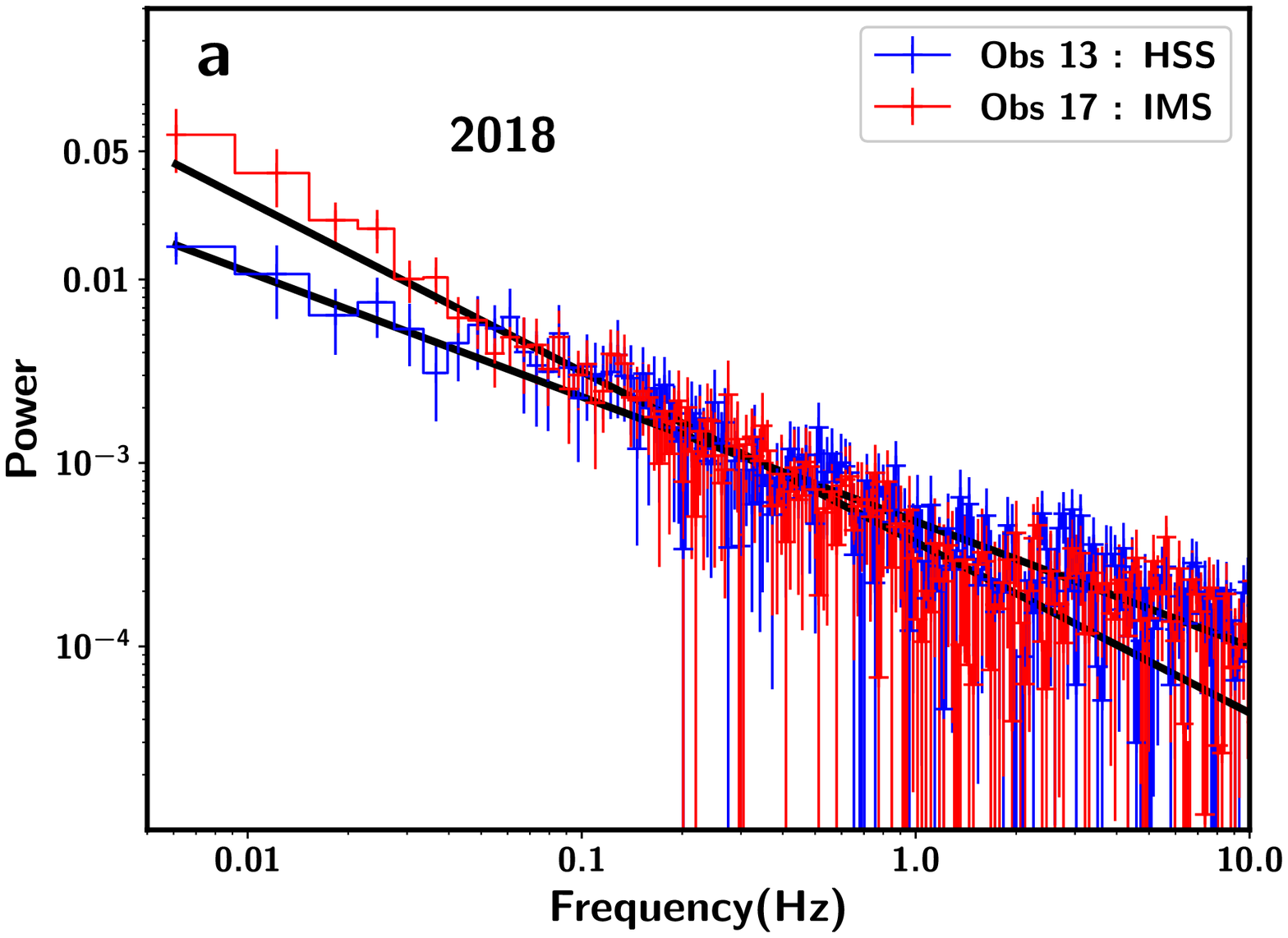}}%
   \end{minipage}%
    \hfill%
    \begin{minipage}{0.5\textwidth}%
    \subfloat[b][\label{fig:rms_var}]{\includegraphics[scale=0.45]{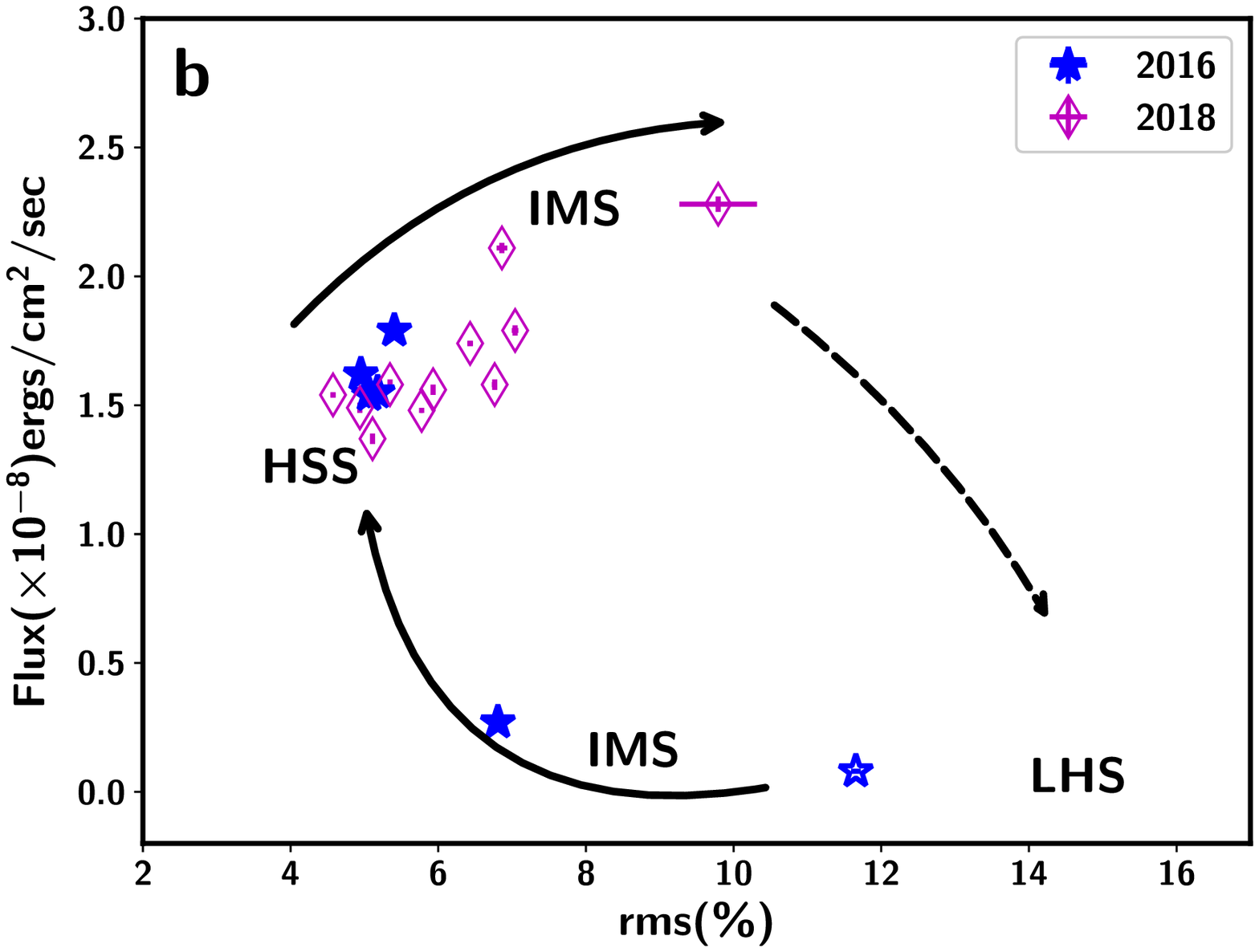}}%
    \end{minipage}%
    \caption{PDS in the energy range \textbf{$3-10$ keV} from 2 different days of 2018 outburst from frequency range $0.005-10$ Hz is plotted on the left. Variation in rms with flux is plotted on the right. The magenta diamonds represent the observations from the 2018 outburst, while the blue filled stars are from 2016. The blue unfilled star corresponds to Obs 2 which is presented for continuity (see text for details). The arrows show the evolution of the rms with spectral states. The dotted arrows show the expected trend in variation of rms with flux.}
\end{figure*}
All the PDS are well modelled by a power law ($AE^{-\beta}$), where A is the normalization and $\beta$ is the index of the power law. No additional component was required. No QPOs were observed during the \textit{AstroSat} campaign of both outbursts. Figure \ref{fig:pds} shows the PDS in the rms space obtained for HSS and IMS of 2018 outburst modelled with a power law for representation. Evolution of the PDS is clearly seen with rms increasing from HSS to intermediate states (Figure \ref{fig:pds}). Variation of rms with flux is plotted in Figure \ref{fig:rms_var}.  The rms decreased from 11.6\% in Obs 2 and 7\% in Obs 3 of 2016 outburst to 4.9\% in the later observations as the source moves to HSS. Total rms decreased from 7\% to 4.9\% as the source moved from IMS to HSS during the 2018 outburst. An increase in rms was again observed during the decay from 5.1\% to 9.8\%. In Figure \ref{fig:rms_var}, we mark the probable transition of the source between states during the rise and the decay. The evolution of rms with different states is discussed in detail in Section \ref{sec:disc}. Solid arrows show the evolution of the source during outburst. Wherever observations are not available, we extrapolate the values based on the observed trend and previously obtained values for other outbursts \citep{Dieters2000,Trudolyubov2001}. This is shown by the dotted arrows in Figure \ref{fig:rms_var}.

\subsection{Mass Estimation}
 \label{sec:mass_est}
Although 4U 1630-472 has undergone multiple outbursts since its discovery which have been studied extensively \citep[and references therein]{Tomsick2005,Seifina2014,Capitanio2015}, a dynamical mass estimate has not been obtained due to high absorption in the direction of the source. Hence, we attempt to constrain the mass of the source using different methods as detailed in the sections below.
\subsubsection{Using the inner disc radius and the bolometric luminosity}
  \label{sec:flux_mass_est}
It is evident that the disc component is prominent during most of the \textit{AstroSat} campaign. Using the apparent inner disc radius ($r_{in}$) obtained from the disc normalization (see Table \ref{tab:pheno_fit}), we can find the actual disc radius ($R_{in}$) using the relation $R_{in}=\xi.\kappa^{2}.r_{in}$ \citep{Kubota1998,Makishima2000}. The inner disc radius is found to vary between 22 to 49 km for inclination varying from $60-70^{\circ}$ \citep{Tomsick1998,Seifina2014,King2014} and distance in the range $10-12$ kpc \citep{Seifina2014,Kalemci2018}. Assuming the disc to be at innermost stable circular orbit (ISCO), the mass of the compact object varies between 2.5 and 5.5 $M_{\odot}$. Assuming a lower mass limit of 3 $M_{\odot}$ for a black hole, the mass range can be quoted as $3-5.5$ $M_{\odot}$.

Secondly, we attempt to constrain the mass of the compact object following the method adopted for IGR J17091-3624 by \cite{Alta2011}. They assumed that the source was emitting at Eddington luminosity at the peak of the $2011-2013$ outburst, where bolometric flux is obtained as \textbf{$F_{Bol}=C_{Bol,Peak}\;F_{2-50\;keV}$} where $C_{Bol,Peak}$ is a bolometric correction factor $\leq$ 3, which is the upper limit as suggested by \cite{Alta2011}. \textit{AstroSat} observed 4U 1630-472 while it underwent a transition from hard to soft state during the rising phase of 2016 outburst and soft to intermediate state during the decay phase of 2018 outburst as a secondary outburst was triggered. We fit the combined spectrum with the phenomenological model and extrapolate in the harder energy range to find the unabsorbed flux from $2-50$ keV. During the soft to intermediate transition in 2018, the maximum flux of 2.1 $\times 10^{-8}$ ergs cm$^{-2}$ s$^{-1}$ was obtained. Considering a representative case where the source emits at Eddington luminosity, we obtain the mass of the black hole as a function of distance for different values of $C_{Bol,Peak}$ (1, 2 \& 3) as the relation between $2-50$ keV flux and bolometric flux is unknown. With $C_{Bol,Peak}=1$, the mass of the compact object lies below 3 $M_{\odot}$ for a distance range of $10-12$ kpc, which is unlikely. Therefore, defining a lower limit of mass as 3 $M_{\odot}$ and $C_{Bol,Peak}$ $>1$ and $\leq$3, mass range is obtained as $3-9$ $M_{\odot}$. However, the mass range so obtained is based on the assumption that the source emits at Eddington luminosity. Hence, in order to understand the accretion dynamics and constrain the mass of the BH independent of the distance and luminosity, we model the broadband spectra for selected observations using the two component advective flow model \citep{Chakrabarti1995,Chakrabarti2006,Iyer2015} in the next section.  
\subsubsection{Modelling with the Two Component Advective Flow}
  \label{sec:tcaf}
\begin{figure*}
    \includegraphics[scale=0.45]{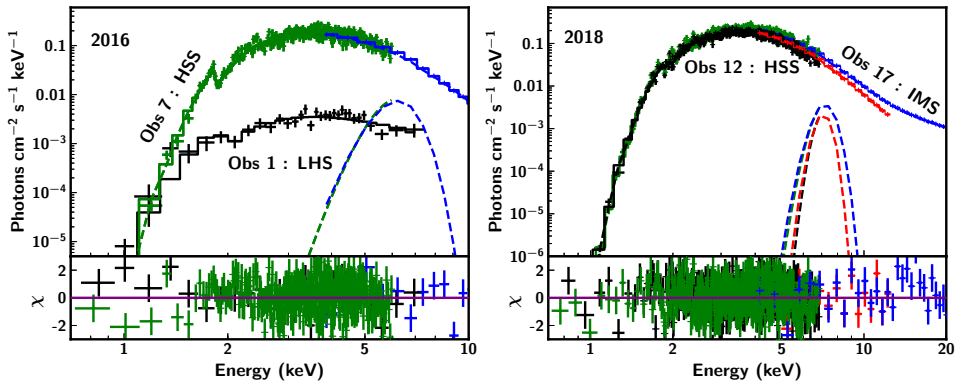}
    \caption{Broadband spectra from LHS and HSS of 2016 outburst fit with two component flow model is shown in the left panel. The right panel shows the spectra from HSS and IMS of 2018 outburst. The residuals are plotted in the bottom panel for both observations. Black and green represent \textit{SXT} observations whereas red and blue correspond to \textit{LAXPC} data.}
     \label{fig:tcaf_spec}
\end{figure*}
\begin{table*}
	\centering
	\begin{threeparttable}
	\caption{Spectral fit parameters obtained using the two component flow model.}
	   \label{tab:tcaf_fit}
	  \begin{tabular}{c|c|c|c|c|c} 
		\hline
		Year & \multicolumn{3}{c}{2016} & \multicolumn{2}{c}{2018} \\
		\hline
		 & LHS & IMS & HSS & HSS & IMS\tnote{c} \\
		MJD & 57627.05 & 57627.69 & 57663.03 & 58344.18 & 58372.16\\
		\hline
		$N_{H}$ (at./cm$^{2}$) & 5.1 $\pm$ 0.5 & 7.8 $\pm$ 0.2 & 6.8 $\pm$ 0.2 & 7.1 $\pm$ 0.1 & 7.3 $\pm$ 0.1\\
		Mass ($M_{\odot}$) & 7 $\pm$ 1 & 8.5 $\pm$ 0.2 & 7.2 $\pm$ 0.2 & 7.3 $\pm$ 0.2 & 7.7 $\pm$ 0.2\\
		Shock Location ($r_{g}$) & 27$^p$ & 27 $\pm$ 1 & 6.0 $\pm$ 0.3 & 8.1 $\pm$ 0.1 & 8.9 $\pm$ 0.1\\
		Halo rate ($\dot m_{Edd}$) & 0.03 $\pm$ 0.01 & 0.04 $\pm$ 0.01 & $0.11^{g}$ & $0.12^{g}$ & $0.2^{g}$\\
		Disc rate ($\dot m_{Edd}$) & 0.06 $\pm$ 0.03 & 0.55 $\pm$ 0.04 & 5.24 $\pm$ 0.09 & 4.7 $\pm$ 0.1 & 5.8 $\pm$ 0.1\\
		$\chi^{2}/dof$ & 1.44 (36./25) & 1.39 (193/152) & 1.27 (416/328) & 1.05(388/370) & 1.11 (479/431)\\
		\hline
      \end{tabular}
      \begin{tablenotes}
	   \item[c]IMS observed during the secondary peak
	   \item[p]parameter frozen to the next observation
	   \item[g]Unable to constrain errors as $\dot m_{h}$ plays minimal role in shaping the spectra
	  \end{tablenotes}
	 \end{threeparttable} 
\end{table*}
We apply the two component flow model to the \textit{SXT} spectrum in the energy range $0.7-6$ keV for Obs 1 and broadband energy spectra obtained using \textit{SXT} and \textit{LAXPC} for the rest of the observations in the energy range as specified in Table \ref{tab:app1}. This model has been implemented as a table model in \texttt{XSPEC} \citep{Iyer2015} and applied in several previous works \citep{Nandi2018,Radhika2018,Sreehari2019}. The two component accretion flow model assumes a geometrically thin accretion disc comprising of a Keplerian flow of matter at the equatorial plane, flanked by sub-Keplerian flow of matter on either sides \citep{Chakrabarti1995,Chakrabarti2006}. The radial flow in the Keplerian disc is always subsonic whereas the sub-Keplerian flow is subsonic at the outer edge and becomes supersonic close to the black hole horizon. The supersonic sub-Keplerian flow encounters a centrifugal barrier on its inward journey towards the black hole that causes a shock transition. Here the flow becomes subsonic, with the flow bulk kinetic energy getting converted to thermal energy \citep{Das2001}. This region beyond the shock is called a post-shock corona (hereafter PSC) which is generally hot and dense. This model has two spectral components, namely the Keplerian disc which supplies the multi-colour blackbody photons of which a small fraction are inverse-Comptonized by the hot electrons in the PSC producing the high energy X-ray power law photons \citep{Chakrabarti1995,Chakrabarti2006}. Unlike the free normalization in the power-law component used in phenomenological spectral modelling, the normalization of the high energy component in the two-component model is decided by the geometry of the PSC. This model has five parameters, namely, mass of the BH ($M$), accretion rates of the Keplerian flow ($\dot m_d$) and sub-Keplerian flow ($\dot m_h$), size of the PSC ($x_s$) and a free normalization. We have expressed $M$ in terms of solar mass ($M_{\odot}$), accretion rates in terms of Eddington rate ($\dot m_{Edd} = 1.4 \times 10^{18}$ $M$ gm s$^{-1}$ with 10\% efficiency) and radial distance in units of $r_g = 2GM/c^2$, where G is gravitational constant and c is speed of light in vacuum. 

We have done the spectral modelling of 4U 1630-472 using two component accretion flow model for a few selected observations taken during the 2016 and 2018 outbursts. The source evolved from hard to soft state during the rise of the 2016 outburst, of which \textit{AstroSat} observations exist for the `canonical' hard and soft states, whereas observations from the soft state and the intermediate state are considered for the 2018 outburst (see Table \ref{tab:pheno_fit}, Figures \ref{fig:maxi_fig} and \ref{fig:rms_var}). In the \texttt{XSPEC} implemented table model, the blackbody spectrum is calculated using the effective temperature ($T_{eff}$) of the Keplerian disc. In the present work, we have done a custom implementation of the two-component model. Here, we have calculated the multi-colour blackbody spectrum using a colour temperature instead of effective temperature \citep{Shimura1995} as,
\begin{center}
$F_{\nu}=\pi B_{\nu} (fT_{eff})/f^4$,
\end{center}
where $F_{\nu}$ is the local flux of radiation, $B_{\nu}$ is the Planck function and $f$ is called the hardening factor. This modification is required, particularly for high accretion rates due to the electron scattering at the inner part of the disc. We find  that $f$ varies from $1.88-1.92$ for the spectral fitting during HSS whereas in the LHS the effect of spectral hardening is less obvious.

We have used interstellar absorption model \textit{phabs} along with two component accretion flow model for spectral modelling. An occasional requirement of Fe line represented by \textit{gaussian} is also used. In Figure \ref{fig:tcaf_spec}, we have presented the broadband spectral modelling of combined \textit{SXT} and \textit{LAXPC} data using two component model for 2016 and 2018 outbursts in the left and the right panels separately. As \textit{LAXPC} data could not be used, only the \textit{SXT} spectrum is modelled for the LHS of 2016 outburst. The shock location could not be constrained and hence it was frozen to the value obtained in the next observation which was only a few hours away. For the other observations, we see that the model could fit the spectral data well with reasonably good statistics. The residual variation of each fit is well within 2$\sigma$ as shown in bottom panels of Figure \ref{fig:tcaf_spec}. The model fitted parameters along with the fitting statistics are presented in Table \ref{tab:tcaf_fit}. We observe a high absorption column density as the source is located towards Galactic center. The column density is particularly high in HSS, possibly indicating existence of disc wind \citep{King2012}. As seen from Table \ref{tab:pheno_fit}, the absorption column density values obtained using phenomenological model lie in a similar range. The model fitted parameters for HSS states in both 2016 and 2018 outbursts are consistent as summarized in Table \ref{tab:tcaf_fit}. In general, the Keplerian disc accretion rate is high in the HSS and the inner edge of the disc is closer to the BH whereas in the LHS the size of the PSC is larger with a smaller value of disc accretion rate ($\dot m_d$). Thus the relative balance between these two parameters (accretion rates) explain the state transition observed in 2016 outburst. However, spectral data of 2018 outburst does not show significant spectral evolution (see Table \ref{tab:pheno_fit}), and also the same is observed in the two component model parameters as mentioned in Table \ref{tab:tcaf_fit}. The errors are estimated using the \textit{migrad} method from the \textit{minuit} library \citep{James2004}. All the errors are quoted at 68\% confidence level unless stated otherwise.

Using our model formalism, we can estimate the distance to the source from the model normalization ($N=cos(i)/d^2$, where $d$ is the distance in 10 kpc) of spectral fitting provided the inclination of the system is known. The model normalization may also be affected due to X-ray contribution from the jet/outflow in the LHS and the current version of the model considers the contribution only from the accretion disc. Hence, we particularly consider the HSS observations where the contribution from the jet/outflow is minimum to constrain the distance to the source. The model normalization provides a distance between $12 - 14$ kpc for $i=60-70^{\circ}$. Also, in our spectral modelling, mass of the source is a free parameter and estimates are consistent for both outbursts. As summarized in Table \ref{tab:tcaf_fit}, the mass of the source lies between $6-8.7$ $M_{\odot}$, which is a better constraint on the mass estimate compared to the one obtained in Section \ref{sec:flux_mass_est}. 
\section{Discussion}
 \label{sec:disc}
\subsection{A `mini' sneak peak at the `super'-outburst}
\begin{figure*}
	\includegraphics[scale=0.5]{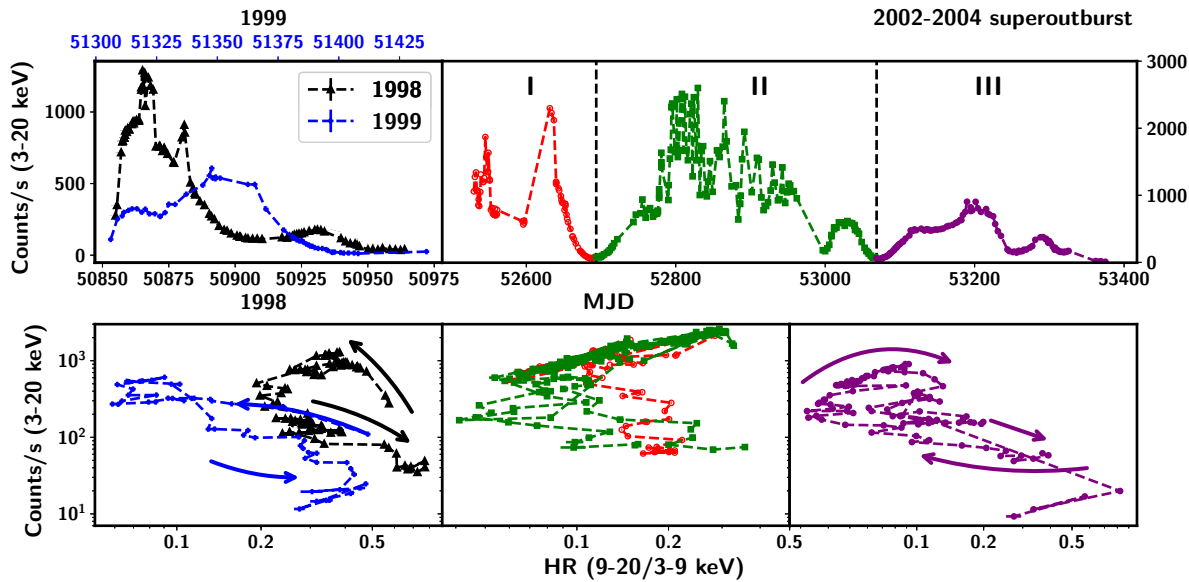}
    \caption{The top panels show the outburst profiles for 1998, 1999 and 2002-2004 outbursts. The bottom panels show the corresponding HIDs with data from \textit{RXTE/PCA} (obtained on request, for details see \protect\cite{Tomsick2005}). The `super'-outburst is divided into 3 regions based on the variation in flux and HR values. The circular head is present in the HID for 1998, 1999 outbursts and the last part of 2002-2004 outburst although it is more defined in the 1999 outburst. The arrows show the direction of evolution in each case. No clear evolution is seen in regions I and II.}
      \label{fig:rxte_hid}
\end{figure*}
We perform detailed studies of the 2016 and 2018 outbursts of 4U  1630-472 using \textit{AstroSat} and \textit{MAXI} data. Both outbursts seem to follow a similar profile, except for a smaller secondary peak seen in the 2018 outburst, before the source decays into quiescence (see Figure \ref{fig:maxi_fig}). During 2016 outburst, the source undergoes a quick transition from hard to soft state through an intermediate state. The IMS of decay phase is longer than that in the rising phase. The source begins at the lower right of the HID in the LHS and moves to the upper left towards HSS. Slight variation in flux and hardness ratio is seen hinting at a transition to an intermediate state. This is represented by the circular head in Figure \ref{fig:maxi_hid}. Then it moves back to the lower part of the plot completing an elongated `c'. The source follows the path traced out by the dotted lines in Figure \ref{fig:maxi_hid} during the 2018 outburst too. However, the soft state in the 2018 outburst has slightly higher HR values in comparison with the 2016 outburst. Similar trend was seen during the 2010 outburst where a secondary peak was observed during the decay phase at similar flux \citep{Capitanio2015}. The HID of these outbursts does not follow the typical `q'-profile (Figure \ref{fig:maxi_hid}) which is in agreement with previous works \citep{Tomsick2005,Capitanio2015}. However, due to the large error bars seen in the HID, the path traced out in the upper right of the HID is difficult to discern clearly. Hence, we use \textit{RXTE/PCA} data obtained during the 1998 and 1999 outburst (see Figure \ref{fig:rxte_hid}, left panels) to further confirm the profile observed in the HID of both outbursts. We consider these two outbursts as an example for classical `mini'-outburst (1999) and a short duration outburst with radio detection (1998). 

The hardness ratio is defined as ratio of counts in $9-20$ keV band to the counts in $3-9$ keV band. This is plotted against the total counts in $3-20$ keV band in the left panels of Figure \ref{fig:rxte_hid}. The variation of the HR from 0.6 to $<$ 0.1 takes place in a few days during the 1999 ouburst, while the transition is more gradual in the 1998 outburst. This is also seen in the 2016 and 2018 outbursts where the transition from hard to soft state takes place very fast (see Figure \ref{fig:maxi_fig}).  A circular head is seen in both HIDs (Figure \ref{fig:rxte_hid}) which could correspond to the intermediate region observed in 2016 and 2018 outbursts. Finally the source goes to the hard state through an intermediate state during the decay (see Figure \ref{fig:rxte_hid}). This is similar to what is observed by \cite{Capitanio2015} for the 2006, 2008 and 2010 outbursts. They reported that there is no evidence of an initial transition from a hard state to a soft state in the 2006 and 2008 outbursts. They suggest that the transition happened very quickly or has not occurred at all.  However, the first \textit{RXTE/PCA} observations of the source in 2006 was carried out $\sim$ 15 d after an increase in \textit{ASM} flux was reported and that for 2008 was done after 6 d of the reported increase in flux. We show here for the first time, that the hard to soft state transition occurs in a short time and has probably not been picked up for previous `mini'-outbursts. \textit{AstroSat} observed the source on the first day itself of the 2016 outburst which shows that a transition from hard to soft state occurred within a few hours (see Figure \ref{fig:maxi_fig}). It is clear from Figures \ref{fig:maxi_fig} and \ref{fig:maxi_hid}, that transition between different states has taken place in both outbursts. Although the HID for 1998 outburst is harder and the circular head is much smaller in the HID, we propose that the evolution of the source during the outbursts of shorter duration follows the `c'-shaped pattern as traced out in Figure \ref{fig:maxi_hid}. Also, the transition from soft to hard state during the decay phase  was seen to occur at slightly lower luminosities  than the rising phase hard to soft transition. Lower values of luminosity during soft to hard transition was also observed during the decay of the 2010 outburst \citep{Tomsick2014}. This is further corroborated by the HID obtained using \textit{RXTE/PCA}  data for the 1998 and 1999 outbursts of this source (Figure \ref{fig:rxte_hid}).

We also look into the outburst profile and HID of the 2002-2004 `super'-outburst to compare it with that of the `mini'-outburst. We divide the 2002-2004 lightcurve into 3 regions with each region comprising of a rise, peak and decay to lower flux. This is shown in the top right panel of Figure \ref{fig:rxte_hid}. It is found that the flux peaks in regions I and III occur with a time interval of $\sim$ 600 days between them which is close to the recurrence period of the `mini'-outbursts \citep{Capitanio2015}. Encouraged by this, we plot the HIDs corresponding to these  regions in the middle (region I and II) and right panels (region III) at the bottom. There does not seem to be any significant pattern in the HID for regions I and II, with variation noticed in the HR only at higher flux. The outburst profile and HID of region III appear to be similar to that of the `mini'-outbursts with a well defined pattern seen in the HID (rightmost panels of Figure \ref{fig:rxte_hid}). The duration of this `mini'-outburst embedded within the `super'-outburst is almost twice the regular duration of $150-200$ days, extending to 311 days. Also, the small circular head in the HID of `mini'-outbursts is greatly elongated and more defined in this segment of the `super'-outburst. Similar outburst profile is also seen for the 2012-2014 `super'-outburst observed with \textit{MAXI} although the duration of this `mini'-outburst is $\sim$ 200 days. Unfortunately, lack of complete coverage of the outburst by better resolution instruments prevents us from commenting on the HID for this duration. The variation in flux of a BHB is generally attributed to the variation of mass accretion rate which power the outbursts. The 2002-2004 outburst was considered a standard outburst for 4U 1630-472 where clear state transitions were seen driven by change in mass accretion rate \citep{Tomsick2005}. It was proposed by \cite{Capitanio2015} that the recurrent `mini'-outbursts could be due to periodic perturbations in the system, which trigger this variation and force the disc temperature to increase independent of the disc accretion rate. These periodic perturbations can be attributed to limit cycle of accretion disc ionization instability \citep{Cannizzo1993,Janiuk2011}, a third body orbiting around the binary system in a hierarchical configuration \citep{Parmar1995} or the constant refilling from a companion star \citep{Trudolyubov2001}, all of which have been reviewed by \cite{Capitanio2015}. In this case, the presence of a complete `mini'-outburst towards the end of the `super'-outburst could suggest that the underlying  mechanism for the origin of `super' and `mini'-outbursts could be the same with varying contributions from some external perturbation which is consistently present for all outbursts. Detailed spectro-temporal analysis of these two different types of outbursts for its entire duration could hold the key to further understanding of their physical origins.

\subsection{Spectral and temporal evolution of 2016 and 2018 outbursts}

In light of the inferences obtained from the outburst profiles and HIDs of the source, we attempt to justify the classification of the outburst into different states based on the spectral and temporal evolution. As is clear from Figure \ref{fig:pheno_fit_fig}, the source lies in the hard state at the beginning of the 2016 outburst with $\Gamma$ $\sim$ 1.82 and the absence of a disc component. The disc appears in Obs 3, indicating a quick transition to an intermediate state. Although contamination from IGR source makes it difficult to quote the values of rms during the hard state, the value obtained is in the range $11-13$ \% which is close to the rms expected from BHBs in LHS i.e., $10-30$\%. \cite{Tomsick2005} obtained rms in the hard state as 9.6\% during the $2002-2004$ outburst, where a significant QPO at $\sim$ 5 Hz was also reported. However, lower count rate and lack of significant power above 1 Hz could have possibly contributed to the non-detection of the QPO in this case. Subsequent observations require a prominent disc component and have lower rms ($<$ 5\%) in the 2016 outburst, similar to the rms obtained by \cite{Pahari2018}, suggesting that the source remained in the soft state (see Figure \ref{fig:rms_var}) for the remainder of the \textit{AstroSat} campaign of 2016. This is also accompanied by an increase in flux in the $0.7 - 20$ keV range (see Figure \ref{fig:rms_var}). The 2018 outburst observations of \textit{AstroSat} seem to begin in the intermediate state with total integrated rms close to $\sim$ 7\% (see Figure \ref{fig:rms_var}). For spectra which could extend beyond 10 keV, the power law index increases till Obs 14, after which a sharp decline is seen till the end of the \textit{AstroSat} campaign (see Table \ref{tab:pheno_fit}).  The corresponding rms values also show a similar trend, with an initial decrease towards the rms obtained for the soft state of 2016 and then show a rise to $\sim$ 10\% towards the end of the 2018 outburst as the source brightens in the hard energy band (see Figure \ref{fig:pheno_fit_fig}) where the secondary peak is observed (see Figure \ref{fig:maxi_fig}). Based on previous studies \citep{Dieters2000,Trudolyubov2001} and the observed trend during these two outbursts of the source, we can expect the rms in the hard state to reach a value of $\sim$ 14\%, which is shown by the dotted arrow in Figure \ref{fig:rms_var}.  

As is evident from Figure \ref{fig:pheno_fit_fig} and Table \ref{tab:pheno_fit}, the disc remains prominent for most of the duration of \textit{AstroSat} campaign of 2016 and 2018 outburst. This suggests that the disc may extend close to the black hole, with a patchy corona on the disc surface \citep{Zdziarski1996} for the HSS of 2016 outburst and all the available observations of 2018 outburst. Not much variation is seen in the luminosity and temperature of source throughout the \textit{AstroSat} campaign of 2016 and 2018 outbursts, except at the initial phase of the 2016 outburst. Therefore, a detailed discussion on the presence of different regimes proposed by \cite{Kubota2001,Kubota2004} cannot be undertaken.  Further comments on the geometry of the source cannot be made as application of more complicated models is limited by the contamination of the IGR source.

\subsection{Exploring an alternative perspective using two component flow model}
Although this source has been studied extensively, the mystery of the recurring outbursts and `missing' hard states have prompted varied explanations on the nature and geometry of the source \citep{Parmar1995,Trudolyubov2001,Capitanio2015}. In this section, we explore an alternative model to explain the spectral evolution of the source under the purview of the two component flow model.

We took three and two observations each, representative of different states during the 2016 and 2018 outbursts with clear variation seen in parameters. As detailed in Table \ref{tab:tcaf_fit} and  Section \ref{sec:tcaf}, an increase in the disc accretion rate is seen as the source transits from LHS to HSS. The size of the PSC is also larger in the IMS (see Table \ref{tab:tcaf_fit}). The larger value of the shock location during LHS and IMS of 2016 outburst relative to the other observations is consistent with the fact that the disc resides away from the black hole. Disc accretion rates are higher in all observations of 2018 outburst where the disc component is prominent. However, there is also a slight increase in the halo accretion rate, accompanied by an increase in the shock location, suggesting an impending transition to a harder state. Although, it is to be noted that the errors on the halo accretion rates could not be constrained as the role of halo contribution is minimal ($\dot m_{h} << \dot m_{d}$) for the softer states. We find that the disc accretion rate is very high ($\sim$ 5 $\dot m_{Edd}$) for the soft states in both outbursts where the luminosity is close to Eddington luminosity for a given distance and inclination of the source. This is because the Keplerian disc terminates at $6-9$ $r_{g}$ (see Table \ref{tab:tcaf_fit}) where the accretion efficiency is $\sim$ 2.5\% for a Schwarzschild black hole. We expect this accretion rate to be reduced by a factor of 5 for an extremely rotating black hole. Theoretical studies of the two-component flow on a spinning BH \citep{Mandal2018,Dihingia2020} show that the spectrum becomes relatively harder and is consistent with the reduction of accretion rate. However, inclusion of spin of the BH may alter the mass of the BH slightly as the accretion efficiency is weakly dependent on mass. It should also be stated that the contribution from jet/outflow are not taken into consideration for this application of the model. This could also affect the model normalisation as the current version of the model only considers the X-ray contribution from the accretion disc, as mentioned earlier. Also, an independent inclusion of \textit{gaussian} is required along with the two component flow model because the associated physical process is not self-consistently included in the model currently. These could slightly affect the fit parameters. Following the inclusion of spin of the BH and these modifications, the model will be applied on previous outbursts of the source for comprehensive understanding which will be presented elsewhere.     
\section{Summary and Conclusions}
 \label{sec:conc}
In this paper, we study the 2016 and 2018 outbursts data using \textit{AstroSat} and \textit{MAXI}. Based on the HID and spectral parameters, we attempt to classify the outburst on different states. We also try to obtain a mass estimate using different methods. The results are summarized as follows:
\begin{itemize}
\item{We attempt a global picture of the outburst and evolution of the source 4U 1630-472 for the `mini'-outbursts which last from $150-200$ days based on the 2016, 2018 outbursts as shown in Figures \ref{fig:maxi_fig} \& \ref{fig:maxi_hid}. We also cross-verify it with the evolution of the 1998 and 1999 outbursts. The HID seems to follow a `c'-shaped profile with a circular head at the upper end. A similar profile is seen towards the end of the `super'-outburst of 2002-2004 with an elongated head.}
\item{The spectral and temporal characteristics of the source during the course of the two outbursts are studied. Although no QPOs are observed during any of the observation, we are able to categorize them into LHS, HSS and IMS states, based on the spectral parameters. However, further classification into HIMS and SIMS is not possible due to non-detection of QPOs and absence of high energy spectra from \textit{LAXPC} due to possible contamination from the nearby source IGR J16320-4751.}
\item{Mass of the compact object in the source 4U 1630-472 is estimated by three different methods, i.e., from the inner disc radius as $3-5.5$ $M_{\odot}$, from the bolometric luminosity as $3-9$ $M_{\odot}$ and and from spectral modelling using the two component flow model as $6-8.7$ $M_{\odot}$. Combining these results, we find the BH mass to lie within a range of $3-9$ $M_{\odot}$.}
\item{The two component flow model seems to satisfactorily explain the spectral state transitions while retaining the preferred geometry for hard and soft states. Application of this model to other outbursts will be attempted later to justify its applicability.} 
\end{itemize}
\section*{Acknowledgements}

We thank the anonymous reviewers for their constructive and insightful suggestions which helped to improve the quality of this paper. We also thank the scientific editor for his/her efforts to expedite the review process. This publication uses the data obtained through High Energy Astrophysics Science Archive Research Center (HEASARC) online service, provided by the NASA/Goddard Space Flight Center. This research has made use of data obtained from \textit{AstroSat} mission of Indian Space Research Organisation (ISRO), archived at the Indian Space Science Data Centre (ISSDC). We thank the \textit{SXT} and \textit{LAXPC} POC at TIFR, Mumbai for verifying and releasing the data. We also thank the respective instrument teams for assistance on the data reduction and analysis. The authors thank John Tomsick, University of California, for providing the \textit{RXTE/PCA} data which facilitated the comparison of HIDs of previous outbursts. BEB, VKA, RMC and AN thank DD, PDMSA and Director, URSC for encouragement and continuous support to carry out this research. TK and HMA acknowledge the support of Department of Atomic Energy, Government of India, under project no. 12-R\& D-TFR-5.02-0200.\\
\\
\textit{Facilities : AstroSat, MAXI, Swift, RXTE}

\section*{Data Availability Statement}
Data underlying this article are available at \textit{AstroSat}-ISSDC website (\url{http://astrobrowse.issdc.gov.in/astro_archive/archive}), \textit{MAXI} website (\url{http://maxi.riken.jp/mxondem}) and \textit{Swift} website (\burl{https://swift.gsfc.nasa.gov/results/transients/}). Data used to generate HIDs for 1998-2004 outbursts was provided by John Tomsick, University of California by permission. This data will be shared on request to the corresponding author with permission of John Tomsick.



\bibliographystyle{mnras}
\bibliography{ref_2.bib} 

\begin{thebibliography}{}
\makeatletter
\relax
\def\mn@urlcharsother{\let\do\@makeother \do\$\do\&\do\#\do\^\do\_\do\%\do\~}
\def\mn@doi{\begingroup\mn@urlcharsother \@ifnextchar [ {\mn@doi@}
  {\mn@doi@[]}}
\def\mn@doi@[#1]#2{\def\@tempa{#1}\ifx\@tempa\@empty \href
  {http://dx.doi.org/#2} {doi:#2}\else \href {http://dx.doi.org/#2} {#1}\fi
  \endgroup}
\def\mn@eprint#1#2{\mn@eprint@#1:#2::\@nil}
\def\mn@eprint@arXiv#1{\href {http://arxiv.org/abs/#1} {{\tt arXiv:#1}}}
\def\mn@eprint@dblp#1{\href {http://dblp.uni-trier.de/rec/bibtex/#1.xml}
  {dblp:#1}}
\def\mn@eprint@#1:#2:#3:#4\@nil{\def\@tempa {#1}\def\@tempb {#2}\def\@tempc
  {#3}\ifx \@tempc \@empty \let \@tempc \@tempb \let \@tempb \@tempa \fi \ifx
  \@tempb \@empty \def\@tempb {arXiv}\fi \@ifundefined
  {mn@eprint@\@tempb}{\@tempb:\@tempc}{\expandafter \expandafter \csname
  mn@eprint@\@tempb\endcsname \expandafter{\@tempc}}}

\bibitem[\protect\citeauthoryear{{Abe}, {Fukazawa}, {Kubota}, {Kasama}  \&
  {Makishima}}{{Abe} et~al.}{2005}]{Abe2005}
{Abe} Y.,  {Fukazawa} Y.,  {Kubota} A.,  {Kasama} D.,   {Makishima} K.,  2005,
  \mn@doi [\pasj] {10.1093/pasj/57.4.629}, \href
  {https://ui.adsabs.harvard.edu/abs/2005PASJ...57..629A} {57, 629}

\bibitem[\protect\citeauthoryear{{Agrawal}}{{Agrawal}}{2006}]{Agrawal2006}
{Agrawal} P.~C.,  2006, in {Wilson} A.,  ed.,  ESA Special Publication Vol.
  604, The X-ray Universe 2005. p.~907

\bibitem[\protect\citeauthoryear{{Agrawal} et~al.,}{{Agrawal}
  et~al.}{2017}]{Agrawal2017}
{Agrawal} P.~C.,  et~al., 2017, \mn@doi [Journal of Astrophysics and Astronomy]
  {10.1007/s12036-017-9451-z}, \href
  {https://ui.adsabs.harvard.edu/abs/2017JApA...38...30A} {38, 30}

\bibitem[\protect\citeauthoryear{{Agrawal}, {Nandi}, {Girish}  \&
  {Ramadevi}}{{Agrawal} et~al.}{2018}]{Agrawal2018}
{Agrawal} V.~K.,  {Nandi} A.,  {Girish} V.,   {Ramadevi} M.~C.,  2018, \mn@doi
  [\mnras] {10.1093/mnras/sty1005}, \href
  {https://ui.adsabs.harvard.edu/abs/2018MNRAS.477.5437A} {477, 5437}

\bibitem[\protect\citeauthoryear{{Altamirano} et~al.,}{{Altamirano}
  et~al.}{2011}]{Alta2011}
{Altamirano} D.,  et~al., 2011, \mn@doi [\apjl] {10.1088/2041-8205/742/2/L17},
  \href {https://ui.adsabs.harvard.edu/abs/2011ApJ...742L..17A} {742, L17}

\bibitem[\protect\citeauthoryear{{Antia} et~al.,}{{Antia}
  et~al.}{2017}]{Antia2017}
{Antia} H.~M.,  et~al., 2017, \mn@doi [\apjs] {10.3847/1538-4365/aa7a0e}, \href
  {https://ui.adsabs.harvard.edu/abs/2017ApJS..231...10A} {231, 10}

\bibitem[\protect\citeauthoryear{{Augusteijn}, {Kuulkers}  \& {van
  Kerkwijk}}{{Augusteijn} et~al.}{2001}]{Augusteijn2001}
{Augusteijn} T.,  {Kuulkers} E.,   {van Kerkwijk} M.~H.,  2001, \mn@doi [\aap]
  {10.1051/0004-6361:20010855}, \href
  {https://ui.adsabs.harvard.edu/abs/2001A&A...375..447A} {375, 447}

\bibitem[\protect\citeauthoryear{{Belloni}}{{Belloni}}{2010}]{Belloni2010}
{Belloni} T.~M.,  2010, {States and Transitions in Black Hole Binaries}.
p.~53, \mn@doi{10.1007/978-3-540-76937-8_3}

\bibitem[\protect\citeauthoryear{{Belloni}, {Homan}, {Casella}, {van der Klis},
  {Nespoli}, {Lewin}, {Miller}  \& {M{\'e}ndez}}{{Belloni}
  et~al.}{2005}]{Belloni2005}
{Belloni} T.,  {Homan} J.,  {Casella} P.,  {van der Klis} M.,  {Nespoli} E.,
  {Lewin} W.~H.~G.,  {Miller} J.~M.,   {M{\'e}ndez} M.,  2005, \mn@doi [\aap]
  {10.1051/0004-6361:20042457}, \href
  {https://ui.adsabs.harvard.edu/abs/2005A&A...440..207B} {440, 207}

\bibitem[\protect\citeauthoryear{{Bildsten} et~al.,}{{Bildsten}
  et~al.}{1997}]{Bildsten1997}
{Bildsten} L.,  et~al., 1997, \mn@doi [\apjs] {10.1086/313060}, \href
  {https://ui.adsabs.harvard.edu/abs/1997ApJS..113..367B} {113, 367}

\bibitem[\protect\citeauthoryear{{Cannizzo}}{{Cannizzo}}{1993}]{Cannizzo1993}
{Cannizzo} J.~K.,  1993, \mn@doi [\apj] {10.1086/173486}, \href
  {https://ui.adsabs.harvard.edu/abs/1993ApJ...419..318C} {419, 318}

\bibitem[\protect\citeauthoryear{{Capitanio}, {Campana}, {De Cesare}  \&
  {Ferrigno}}{{Capitanio} et~al.}{2015}]{Capitanio2015}
{Capitanio} F.,  {Campana} R.,  {De Cesare} G.,   {Ferrigno} C.,  2015, \mn@doi
  [\mnras] {10.1093/mnras/stv687}, \href
  {https://ui.adsabs.harvard.edu/abs/2015MNRAS.450.3840C} {450, 3840}

\bibitem[\protect\citeauthoryear{{Casella}, {Belloni}  \& {Stella}}{{Casella}
  et~al.}{2005}]{Casella2005}
{Casella} P.,  {Belloni} T.,   {Stella} L.,  2005, \mn@doi [\apj]
  {10.1086/431174}, \href
  {https://ui.adsabs.harvard.edu/abs/2005ApJ...629..403C} {629, 403}

\bibitem[\protect\citeauthoryear{{Chakrabarti} \& {Mandal}}{{Chakrabarti} \&
  {Mandal}}{2006}]{Chakrabarti2006}
{Chakrabarti} S.~K.,  {Mandal} S.,  2006, \mn@doi [\apjl] {10.1086/504319},
  \href {https://ui.adsabs.harvard.edu/abs/2006ApJ...642L..49C} {642, L49}

\bibitem[\protect\citeauthoryear{{Chakrabarti} \& {Titarchuk}}{{Chakrabarti} \&
  {Titarchuk}}{1995}]{Chakrabarti1995}
{Chakrabarti} S.,  {Titarchuk} L.~G.,  1995, \mn@doi [\apj] {10.1086/176610},
  \href {http://adsabs.harvard.edu/abs/1995ApJ...455..623C} {455, 623}

\bibitem[\protect\citeauthoryear{{Corral-Santana}, {Casares},
  {Mu{\~n}oz-Darias}, {Bauer}, {Mart{\'\i}nez-Pais}  \&
  {Russell}}{{Corral-Santana} et~al.}{2016}]{Corral2016}
{Corral-Santana} J.~M.,  {Casares} J.,  {Mu{\~n}oz-Darias} T.,  {Bauer} F.~E.,
  {Mart{\'\i}nez-Pais} I.~G.,   {Russell} D.~M.,  2016, \mn@doi [\aap]
  {10.1051/0004-6361/201527130}, \href
  {https://ui.adsabs.harvard.edu/abs/2016A&A...587A..61C} {587, A61}

\bibitem[\protect\citeauthoryear{{Das}, {Chattopadhyay}  \&
  {Chakrabarti}}{{Das} et~al.}{2001}]{Das2001}
{Das} S.,  {Chattopadhyay} I.,   {Chakrabarti} S. i.~K.,  2001, \mn@doi [\apj]
  {10.1086/321692}, \href
  {https://ui.adsabs.harvard.edu/abs/2001ApJ...557..983D} {557, 983}

\bibitem[\protect\citeauthoryear{{Dieters} et~al.,}{{Dieters}
  et~al.}{2000}]{Dieters2000}
{Dieters} S.~W.,  et~al., 2000, \mn@doi [\apj] {10.1086/309108}, \href
  {https://ui.adsabs.harvard.edu/abs/2000ApJ...538..307D} {538, 307}

\bibitem[\protect\citeauthoryear{Dihingia, Das, Prabhakar  \& Mandal}{Dihingia
  et~al.}{2020}]{Dihingia2020}
Dihingia I.~K.,  Das S.,  Prabhakar G.,   Mandal S.,  2020, \mnras, doi :
  10.1093/mnras/staa1687

\bibitem[\protect\citeauthoryear{{Fender}, {Belloni}  \& {Gallo}}{{Fender}
  et~al.}{2004}]{Fender2004}
{Fender} R.~P.,  {Belloni} T.~M.,   {Gallo} E.,  2004, \mn@doi [\mnras]
  {10.1111/j.1365-2966.2004.08384.x}, \href
  {https://ui.adsabs.harvard.edu/abs/2004MNRAS.355.1105F} {355, 1105}

\bibitem[\protect\citeauthoryear{{Fender}, {Homan}  \& {Belloni}}{{Fender}
  et~al.}{2009}]{Fender2009}
{Fender} R.~P.,  {Homan} J.,   {Belloni} T.~M.,  2009, \mn@doi [\mnras]
  {10.1111/j.1365-2966.2009.14841.x}, \href
  {https://ui.adsabs.harvard.edu/abs/2009MNRAS.396.1370F} {396, 1370}

\bibitem[\protect\citeauthoryear{{Homan} \& {Belloni}}{{Homan} \&
  {Belloni}}{2005}]{Homan2005}
{Homan} J.,  {Belloni} T.,  2005, \mn@doi [\apss] {10.1007/s10509-005-1197-4},
  \href {https://ui.adsabs.harvard.edu/abs/2005Ap&SS.300..107H} {300, 107}

\bibitem[\protect\citeauthoryear{{Homan}, {Wijnands}, {van der Klis},
  {Belloni}, {van Paradijs}, {Klein-Wolt}, {Fender}  \& {M{\'e}ndez}}{{Homan}
  et~al.}{2001}]{Homan2001}
{Homan} J.,  {Wijnands} R.,  {van der Klis} M.,  {Belloni} T.,  {van Paradijs}
  J.,  {Klein-Wolt} M.,  {Fender} R.,   {M{\'e}ndez} M.,  2001, \mn@doi [\apjs]
  {10.1086/318954}, \href
  {https://ui.adsabs.harvard.edu/abs/2001ApJS..132..377H} {132, 377}

\bibitem[\protect\citeauthoryear{{Iyer}, {Nandi}  \& {Mandal}}{{Iyer}
  et~al.}{2015}]{Iyer2015}
{Iyer} N.,  {Nandi} A.,   {Mandal} S.,  2015, \mn@doi [\apj]
  {10.1088/0004-637X/807/1/108}, \href
  {https://ui.adsabs.harvard.edu/abs/2015ApJ...807..108I} {807, 108}

\bibitem[\protect\citeauthoryear{James \& Winkler}{James \&
  Winkler}{2004}]{James2004}
James F.,  Winkler M.,  2004, MINUIT User's Guide,
  http://seal.web.cern.ch/seal/documents/minuit/mnusers-guide.pdf

\bibitem[\protect\citeauthoryear{{Janiuk} \& {Czerny}}{{Janiuk} \&
  {Czerny}}{2011}]{Janiuk2011}
{Janiuk} A.,  {Czerny} B.,  2011, \mn@doi [\mnras]
  {10.1111/j.1365-2966.2011.18544.x}, \href
  {https://ui.adsabs.harvard.edu/abs/2011MNRAS.414.2186J} {414, 2186}

\bibitem[\protect\citeauthoryear{{Jones}, {Forman}, {Tananbaum}  \&
  {Turner}}{{Jones} et~al.}{1976}]{Jones1976}
{Jones} C.,  {Forman} W.,  {Tananbaum} H.,   {Turner} M.~J.~L.,  1976, \mn@doi
  [\apj] {10.1086/182291}, \href
  {https://ui.adsabs.harvard.edu/abs/1976ApJ...210L...9J} {210, L9}

\bibitem[\protect\citeauthoryear{{Kalemci}, {Maccarone}  \&
  {Tomsick}}{{Kalemci} et~al.}{2018}]{Kalemci2018}
{Kalemci} E.,  {Maccarone} T.~J.,   {Tomsick} J.~A.,  2018, \mn@doi [\apj]
  {10.3847/1538-4357/aabcd3}, \href
  {https://ui.adsabs.harvard.edu/abs/2018ApJ...859...88K} {859, 88}

\bibitem[\protect\citeauthoryear{{King}, {Miller}  \& {Raymond}}{{King}
  et~al.}{2012}]{King2012}
{King} A.~L.,  {Miller} J.~M.,   {Raymond} J.,  2012, \mn@doi [\apj]
  {10.1088/0004-637X/746/1/2}, \href
  {https://ui.adsabs.harvard.edu/abs/2012ApJ...746....2K} {746, 2}

\bibitem[\protect\citeauthoryear{{King} et~al.,}{{King}
  et~al.}{2014}]{King2014}
{King} A.~L.,  et~al., 2014, \mn@doi [\apj] {10.1088/2041-8205/784/1/L2}, \href
  {https://ui.adsabs.harvard.edu/abs/2014ApJ...784L...2K} {784, L2}

\bibitem[\protect\citeauthoryear{{Kubota} \& {Makishima}}{{Kubota} \&
  {Makishima}}{2004}]{Kubota2004}
{Kubota} A.,  {Makishima} K.,  2004, \mn@doi [\apj] {10.1086/380433}, \href
  {https://ui.adsabs.harvard.edu/abs/2004ApJ...601..428K} {601, 428}

\bibitem[\protect\citeauthoryear{{Kubota}, {Tanaka}, {Makishima}, {Ueda},
  {Dotani}, {Inoue}  \& {Yamaoka}}{{Kubota} et~al.}{1998}]{Kubota1998}
{Kubota} A.,  {Tanaka} Y.,  {Makishima} K.,  {Ueda} Y.,  {Dotani} T.,  {Inoue}
  H.,   {Yamaoka} K.,  1998, \mn@doi [\pasj] {10.1093/pasj/50.6.667}, \href
  {https://ui.adsabs.harvard.edu/abs/1998PASJ...50..667K} {50, 667}

\bibitem[\protect\citeauthoryear{{Kubota}, {Makishima}  \& {Ebisawa}}{{Kubota}
  et~al.}{2001}]{Kubota2001}
{Kubota} A.,  {Makishima} K.,   {Ebisawa} K.,  2001, \mn@doi [\apj]
  {10.1086/324377}, \href
  {https://ui.adsabs.harvard.edu/abs/2001ApJ...560L.147K} {560, L147}

\bibitem[\protect\citeauthoryear{{Kuulkers}, {Wijnands}, {Belloni},
  {M{\'e}ndez}, {van der Klis}  \& {van Paradijs}}{{Kuulkers}
  et~al.}{1998}]{Kuulkers1998}
{Kuulkers} E.,  {Wijnands} R.,  {Belloni} T.,  {M{\'e}ndez} M.,  {van der Klis}
  M.,   {van Paradijs} J.,  1998, \mn@doi [\apj] {10.1086/305248}, \href
  {https://ui.adsabs.harvard.edu/abs/1998ApJ...494..753K} {494, 753}

\bibitem[\protect\citeauthoryear{{Lewin} \& {Livingston}}{{Lewin} \&
  {Livingston}}{1995}]{Lewin1995}
{Lewin} W.~H.~G.,  {Livingston} W.,  1995, Journal of the British Astronomical
  Association, \href {https://ui.adsabs.harvard.edu/abs/1995JBAA..105R.284L}
  {105, 284}

\bibitem[\protect\citeauthoryear{{Lewin} \& {van der Klis}}{{Lewin} \& {van der
  Klis}}{2006}]{Klis2006}
{Lewin} W. H.~G.,  {van der Klis} M.,  2006, {Compact Stellar X-ray Sources}.
 Vol. 39

\bibitem[\protect\citeauthoryear{{Maccarone} \& {Coppi}}{{Maccarone} \&
  {Coppi}}{2003}]{Maccarone2003}
{Maccarone} T.~J.,  {Coppi} P.~S.,  2003, \mn@doi [\mnras]
  {10.1046/j.1365-8711.2003.06040.x}, \href
  {https://ui.adsabs.harvard.edu/abs/2003MNRAS.338..189M} {338, 189}

\bibitem[\protect\citeauthoryear{{Makishima}, {Maejima}, {Mitsuda}, {Bradt},
  {Remillard}, {Tuohy}, {Hoshi}  \& {Nakagawa}}{{Makishima}
  et~al.}{1986}]{Makishima1986}
{Makishima} K.,  {Maejima} Y.,  {Mitsuda} K.,  {Bradt} H.~V.,  {Remillard}
  R.~A.,  {Tuohy} I.~R.,  {Hoshi} R.,   {Nakagawa} M.,  1986, \mn@doi [\apj]
  {10.1086/164534}, \href
  {https://ui.adsabs.harvard.edu/abs/1986ApJ...308..635M} {308, 635}

\bibitem[\protect\citeauthoryear{{Makishima} et~al.,}{{Makishima}
  et~al.}{2000}]{Makishima2000}
{Makishima} K.,  et~al., 2000, \mn@doi [\apj] {10.1086/308868}, \href
  {https://ui.adsabs.harvard.edu/abs/2000ApJ...535..632M} {535, 632}

\bibitem[\protect\citeauthoryear{{Mandal} \& {Mondal}}{{Mandal} \&
  {Mondal}}{2018}]{Mandal2018}
{Mandal} S.,  {Mondal} S.,  2018, \mn@doi [Journal of Astrophysics and
  Astronomy] {10.1007/s12036-017-9509-y}, \href
  {https://ui.adsabs.harvard.edu/abs/2018JApA...39...19M} {39, 19}

\bibitem[\protect\citeauthoryear{{Misra} et~al.,}{{Misra}
  et~al.}{2017}]{Misra2017}
{Misra} R.,  et~al., 2017, \mn@doi [\apj] {10.3847/1538-4357/835/2/195}, \href
  {https://ui.adsabs.harvard.edu/abs/2017ApJ...835..195M} {835, 195}

\bibitem[\protect\citeauthoryear{{Mitsuda} et~al.,}{{Mitsuda}
  et~al.}{1984}]{Mitsuda1984}
{Mitsuda} K.,  et~al., 1984, \pasj, \href
  {https://ui.adsabs.harvard.edu/abs/1984PASJ...36..741M} {36, 741}

\bibitem[\protect\citeauthoryear{{Motta}, {Belloni}  \& {Homan}}{{Motta}
  et~al.}{2009}]{Motta2009}
{Motta} S.,  {Belloni} T.,   {Homan} J.,  2009, \mn@doi [\mnras]
  {10.1111/j.1365-2966.2009.15566.x}, \href
  {https://ui.adsabs.harvard.edu/abs/2009MNRAS.400.1603M} {400, 1603}

\bibitem[\protect\citeauthoryear{{Motta}, {Homan}, {Mu{\~n}oz Darias},
  {Casella}, {Belloni}, {Hiemstra}  \& {M{\'e}ndez}}{{Motta}
  et~al.}{2012}]{Motta2012}
{Motta} S.,  {Homan} J.,  {Mu{\~n}oz Darias} T.,  {Casella} P.,  {Belloni}
  T.~M.,  {Hiemstra} B.,   {M{\'e}ndez} M.,  2012, \mn@doi [\mnras]
  {10.1111/j.1365-2966.2012.22037.x}, \href
  {https://ui.adsabs.harvard.edu/abs/2012MNRAS.427..595M} {427, 595}

\bibitem[\protect\citeauthoryear{{Nandi}, {Debnath}, {Mandal}  \&
  {Chakrabarti}}{{Nandi} et~al.}{2012}]{Nandi2012}
{Nandi} A.,  {Debnath} D.,  {Mandal} S.,   {Chakrabarti} S.~K.,  2012, \mn@doi
  [\aap] {10.1051/0004-6361/201117844}, \href
  {https://ui.adsabs.harvard.edu/abs/2012A&A...542A..56N} {542, A56}

\bibitem[\protect\citeauthoryear{{Nandi} et~al.,}{{Nandi}
  et~al.}{2018}]{Nandi2018}
{Nandi} A.,  et~al., 2018, \mn@doi [\apss] {10.1007/s10509-018-3314-1}, \href
  {https://ui.adsabs.harvard.edu/abs/2018Ap&SS.363...90N} {363, 90}

\bibitem[\protect\citeauthoryear{{Narayan} \& {Yi}}{{Narayan} \&
  {Yi}}{1995}]{Narayan1995}
{Narayan} R.,  {Yi} I.,  1995, \mn@doi [\apj] {10.1086/176343}, \href
  {https://ui.adsabs.harvard.edu/abs/1995ApJ...452..710N} {452, 710}

\bibitem[\protect\citeauthoryear{{Pahari} et~al.,}{{Pahari}
  et~al.}{2018}]{Pahari2018}
{Pahari} M.,  et~al., 2018, \mn@doi [\apj] {10.3847/1538-4357/aae53b}, \href
  {https://ui.adsabs.harvard.edu/abs/2018ApJ...867...86P} {867, 86}

\bibitem[\protect\citeauthoryear{{Parmar}, {Angelini}  \& {White}}{{Parmar}
  et~al.}{1995}]{Parmar1995}
{Parmar} A.~N.,  {Angelini} L.,   {White} N.~E.,  1995, \mn@doi [\apj]
  {10.1086/309730}, \href
  {https://ui.adsabs.harvard.edu/abs/1995ApJ...452L.129P} {452, L129}

\bibitem[\protect\citeauthoryear{{Parmar}, {Williams}, {Kuulkers}, {Angelini}
  \& {White}}{{Parmar} et~al.}{1997}]{Parmar1997}
{Parmar} A.~N.,  {Williams} O.~R.,  {Kuulkers} E.,  {Angelini} L.,   {White}
  N.~E.,  1997, \aap, \href
  {https://ui.adsabs.harvard.edu/abs/1997A&A...319..855P} {319, 855}

\bibitem[\protect\citeauthoryear{{Priedhorsky}}{{Priedhorsky}}{1986}]{Priedhorsky1986}
{Priedhorsky} W.,  1986, \mn@doi [\apss] {10.1007/BF00644177}, \href
  {https://ui.adsabs.harvard.edu/abs/1986Ap&SS.126...89P} {126, 89}

\bibitem[\protect\citeauthoryear{{Radhika}, {Nandi}, {Agrawal}  \&
  {Seetha}}{{Radhika} et~al.}{2016}]{Radhika2016}
{Radhika} D.,  {Nandi} A.,  {Agrawal} V.~K.,   {Seetha} S.,  2016, \mn@doi
  [\mnras] {10.1093/mnras/stw1239}, \href
  {https://ui.adsabs.harvard.edu/abs/2016MNRAS.460.4403R} {460, 4403}

\bibitem[\protect\citeauthoryear{{Radhika}, {Sreehari}, {Nandi}, {Iyer}  \&
  {Mand al}}{{Radhika} et~al.}{2018}]{Radhika2018}
{Radhika} D.,  {Sreehari} H.,  {Nandi} A.,  {Iyer} N.,   {Mand al} S.,  2018,
  \mn@doi [\apss] {10.1007/s10509-018-3411-1}, \href
  {https://ui.adsabs.harvard.edu/abs/2018Ap&SS.363..189D} {363, 189}

\bibitem[\protect\citeauthoryear{{Ramadevi} et~al.,}{{Ramadevi}
  et~al.}{2017}]{Ramadevi2017}
{Ramadevi} M.~C.,  et~al., 2017, \mn@doi [Journal of Astrophysics and
  Astronomy] {10.1007/s12036-017-9454-9}, \href
  {https://ui.adsabs.harvard.edu/abs/2017JApA...38...32R} {38, 32}

\bibitem[\protect\citeauthoryear{{Remillard} \& {McClintock}}{{Remillard} \&
  {McClintock}}{2006}]{Remillard2006}
{Remillard} R.~A.,  {McClintock} J.~E.,  2006, \mn@doi [\araa]
  {10.1146/annurev.astro.44.051905.092532}, \href
  {https://ui.adsabs.harvard.edu/abs/2006ARA&A..44...49R} {44, 49}

\bibitem[\protect\citeauthoryear{{Rodriguez}, {Tomsick}, {Foschini}, {Walter},
  {Goldwurm}, {Corbel}  \& {Kaaret}}{{Rodriguez} et~al.}{2003}]{Rodriguez2003}
{Rodriguez} J.,  {Tomsick} J.~A.,  {Foschini} L.,  {Walter} R.,  {Goldwurm} A.,
   {Corbel} S.,   {Kaaret} P.,  2003, \mn@doi [\aap]
  {10.1051/0004-6361:20031093}, \href
  {https://ui.adsabs.harvard.edu/abs/2003A&A...407L..41R} {407, L41}

\bibitem[\protect\citeauthoryear{{Rodriguez} et~al.,}{{Rodriguez}
  et~al.}{2006}]{Rodriguez2006}
{Rodriguez} J.,  et~al., 2006, \mn@doi [\mnras]
  {10.1111/j.1365-2966.2005.09855.x}, \href
  {https://ui.adsabs.harvard.edu/abs/2006MNRAS.366..274R} {366, 274}

\bibitem[\protect\citeauthoryear{{Seetha} et~al.,}{{Seetha}
  et~al.}{2006}]{Seetha2006}
{Seetha} S.,  et~al., 2006, \mn@doi [Advances in Space Research]
  {10.1016/j.asr.2005.09.046}, \href
  {https://ui.adsabs.harvard.edu/abs/2006AdSpR..38.2995S} {38, 2995}

\bibitem[\protect\citeauthoryear{{Seifina}, {Titarchuk}  \&
  {Shaposhnikov}}{{Seifina} et~al.}{2014}]{Seifina2014}
{Seifina} E.,  {Titarchuk} L.,   {Shaposhnikov} N.,  2014, \mn@doi [\apj]
  {10.1088/0004-637X/789/1/57}, \href
  {https://ui.adsabs.harvard.edu/abs/2014ApJ...789...57S} {789, 57}

\bibitem[\protect\citeauthoryear{{Shakura} \& {Sunyaev}}{{Shakura} \&
  {Sunyaev}}{1973}]{Shakura1973}
{Shakura} N.~I.,  {Sunyaev} R.~A.,  1973, \aap, \href
  {https://ui.adsabs.harvard.edu/abs/1973A&A....24..337S} {500, 33}

\bibitem[\protect\citeauthoryear{{Shaposhnikov} \& {Titarchuk}}{{Shaposhnikov}
  \& {Titarchuk}}{2009}]{Shaposhnikov2009}
{Shaposhnikov} N.,  {Titarchuk} L.,  2009, \mn@doi [\apj]
  {10.1088/0004-637X/699/1/453}, \href
  {https://ui.adsabs.harvard.edu/abs/2009ApJ...699..453S} {699, 453}

\bibitem[\protect\citeauthoryear{{Share} et~al.,}{{Share}
  et~al.}{1978}]{Kaluzienski1978}
{Share} G.,  et~al., 1978, \iaucirc, \href
  {https://ui.adsabs.harvard.edu/abs/1978IAUC.3197....3S} {3197, 3}

\bibitem[\protect\citeauthoryear{{Shimura} \& {Takahara}}{{Shimura} \&
  {Takahara}}{1995}]{Shimura1995}
{Shimura} T.,  {Takahara} F.,  1995, \mn@doi [\apj] {10.1086/175740}, \href
  {https://ui.adsabs.harvard.edu/abs/1995ApJ...445..780S} {445, 780}

\bibitem[\protect\citeauthoryear{{Singh} et~al.,}{{Singh}
  et~al.}{2014}]{Singh2014}
{Singh} K.~P.,  et~al., 2014, {ASTROSAT mission}.
 Society of Photo-Optical Instrumentation Engineers (SPIE) Conference Series
  Vol. 9144, \mn@doi{10.1117/12.2062667, }

\bibitem[\protect\citeauthoryear{{Singh} et~al.,}{{Singh}
  et~al.}{2017}]{Singh2017}
{Singh} K.~P.,  et~al., 2017, \mn@doi [Journal of Astrophysics and Astronomy]
  {10.1007/s12036-017-9448-7}, \href
  {https://ui.adsabs.harvard.edu/abs/2017JApA...38...29S} {38, 29}

\bibitem[\protect\citeauthoryear{{Smith}, {Edgar}  \& {Shafer}}{{Smith}
  et~al.}{2002}]{Smith2002}
{Smith} R.~K.,  {Edgar} R.~J.,   {Shafer} R.~A.,  2002, \mn@doi [\apj]
  {10.1086/344151}, \href
  {https://ui.adsabs.harvard.edu/abs/2002ApJ...581..562S} {581, 562}

\bibitem[\protect\citeauthoryear{{Sreehari}, {Nandi}, {Radhika}, {Iyer}  \&
  {Mandal}}{{Sreehari} et~al.}{2018}]{Sreehari2018}
{Sreehari} H.,  {Nandi} A.,  {Radhika} D.,  {Iyer} N.,   {Mandal} S.,  2018,
  \mn@doi [Journal of Astrophysics and Astronomy] {10.1007/s12036-018-9510-0},
  \href {https://ui.adsabs.harvard.edu/abs/2018JApA...39....5S} {39, 5}

\bibitem[\protect\citeauthoryear{{Sreehari}, {Ravishankar}, {Iyer}, {Agrawal},
  {Katoch}, {Mandal}  \& {Nand i}}{{Sreehari} et~al.}{2019}]{Sreehari2019}
{Sreehari} H.,  {Ravishankar} B.~T.,  {Iyer} N.,  {Agrawal} V.~K.,  {Katoch}
  T.~B.,  {Mandal} S.,   {Nand i} A.,  2019, \mn@doi [\mnras]
  {10.1093/mnras/stz1327}, \href
  {https://ui.adsabs.harvard.edu/abs/2019MNRAS.487..928S} {487, 928}

\bibitem[\protect\citeauthoryear{{Tanaka} \& {Lewin}}{{Tanaka} \&
  {Lewin}}{1995}]{Tanaka1995}
{Tanaka} Y.,  {Lewin} W.~H.~G.,  1995, \href
  {https://ui.adsabs.harvard.edu/abs/1995xrbi.nasa..126T} {pp 126--174}

\bibitem[\protect\citeauthoryear{{Tandon} et~al.,}{{Tandon}
  et~al.}{2017}]{Tandon2017}
{Tandon} S.~N.,  et~al., 2017, \mn@doi [\aj] {10.3847/1538-3881/aa8451}, \href
  {https://ui.adsabs.harvard.edu/abs/2017AJ....154..128T} {154, 128}

\bibitem[\protect\citeauthoryear{{Tomsick}, {Lapshov}  \& {Kaaret}}{{Tomsick}
  et~al.}{1998}]{Tomsick1998}
{Tomsick} J.~A.,  {Lapshov} I.,   {Kaaret} P.,  1998, \mn@doi [\apj]
  {10.1086/305240}, \href
  {https://ui.adsabs.harvard.edu/abs/1998ApJ...494..747T} {494, 747}

\bibitem[\protect\citeauthoryear{{Tomsick}, {Corbel}, {Goldwurm}  \&
  {Kaaret}}{{Tomsick} et~al.}{2005}]{Tomsick2005}
{Tomsick} J.~A.,  {Corbel} S.,  {Goldwurm} A.,   {Kaaret} P.,  2005, \mn@doi
  [\apj] {10.1086/431896}, \href
  {https://ui.adsabs.harvard.edu/abs/2005ApJ...630..413T} {630, 413}

\bibitem[\protect\citeauthoryear{{Tomsick} et~al.,}{{Tomsick}
  et~al.}{2014}]{Tomsick2014}
{Tomsick} J.~A.,  et~al., 2014, \mn@doi [\apj] {10.1088/0004-637X/780/1/78},
  \href {https://ui.adsabs.harvard.edu/abs/2014ApJ...780...78T} {780, 78}

\bibitem[\protect\citeauthoryear{{Trudolyubov}, {Borozdin}  \&
  {Priedhorsky}}{{Trudolyubov} et~al.}{2001}]{Trudolyubov2001}
{Trudolyubov} S.~P.,  {Borozdin} K.~N.,   {Priedhorsky} W.~C.,  2001, \mn@doi
  [\mnras] {10.1046/j.1365-8711.2001.04073.x}, \href
  {https://ui.adsabs.harvard.edu/abs/2001MNRAS.322..309T} {322, 309}

\bibitem[\protect\citeauthoryear{{Ueda} et~al.,}{{Ueda}
  et~al.}{2010}]{Ueda2010}
{Ueda} Y.,  et~al., 2010, \mn@doi [\apj] {10.1088/0004-637X/713/1/257}, \href
  {https://ui.adsabs.harvard.edu/abs/2010ApJ...713..257U} {713, 257}

\bibitem[\protect\citeauthoryear{{Vadawale} et~al.,}{{Vadawale}
  et~al.}{2016}]{Vadawale2016}
{Vadawale} S.~V.,  et~al., 2016, {In-orbit performance AstroSat CZTI}.
p. 99051G, \mn@doi{10.1117/12.2235373}

\bibitem[\protect\citeauthoryear{{Yadav} et~al.,}{{Yadav}
  et~al.}{2016}]{Yadav2016a}
{Yadav} J.~S.,  et~al., 2016, The Astronomer's Telegram, \href
  {https://ui.adsabs.harvard.edu/abs/2016ATel.9515....1Y} {9515, 1}

\bibitem[\protect\citeauthoryear{{Zdziarski}, {Johnson}  \&
  {Magdziarz}}{{Zdziarski} et~al.}{1996}]{Zdziarski1996}
{Zdziarski} A.~A.,  {Johnson} W.~N.,   {Magdziarz} P.,  1996, \mn@doi [\mnras]
  {10.1093/mnras/283.1.193}, \href
  {https://ui.adsabs.harvard.edu/abs/1996MNRAS.283..193Z} {283, 193}

\bibitem[\protect\citeauthoryear{{Zhang}, {Jahoda}, {Swank}, {Morgan}  \&
  {Giles}}{{Zhang} et~al.}{1995}]{Zhang1995}
{Zhang} W.,  {Jahoda} K.,  {Swank} J.~H.,  {Morgan} E.~H.,   {Giles} A.~B.,
  1995, \mn@doi [\apj] {10.1086/176111}, \href
  {http://adsabs.harvard.edu/abs/1995ApJ...449..930Z} {449, 930}

\makeatother
\end{thebibliography}



\appendix
\counterwithin{figure}{section}
\counterwithin{table}{section}
\section{Pulsation studies of IGR J16320-4751}
 \label{sec:append}
{\small
\begin{table*}
	\centering
	\begin{threeparttable}
	\caption{\small{Summary of detected pulse period along with typical contribution of IGR J16320-4751 for each observation of 2016 and 2018 outburst. The pulse fraction considered nominal value of 20\% \citep{Rodriguez2006} to estimate the contribution of IGR.}}
		   \label{tab:app1}
	  \begin{tabular}{c@{\hspace{2pt}}c@{\hspace{2pt}}c@{\hspace{2pt}}c@{\hspace{2pt}}c@{\hspace{2pt}}c@{\hspace{2pt}}c@{\hspace{2pt}}c} 
		\hline
		Obs & ObsID & Exp. & Pulse Period & Pulse Amplitude & Typical contribution & IGR cts/Total cts & Energy band \\
		 & & (ks) & (s) & (c/s) & from IGR source (c/s) & (Contamination \%) & used (keV)\\ 
		\hline
		 & & & & 2016 & & & \\
		\hline
		3 & 0626 & 10.4 & 1293 $\pm$ 1 & 6 & 15 & 15/144 (10.4) & $3-10$ \\
		4,5 & 0686 & 51 & 1299 $\pm$ 5 & 6 & 15 & 15/715 (2.1) & $3-13$ \\
		6,7 & 0698 & 41 & 1314 $\pm$ 2 & 8 & 20 & 20/715 (2.8) & $3-13$ \\
		\hline
		 & & & & 2018 & & & \\
		\hline
		8,9 & 2274 & 24 & 1316 $\pm$ 4 & 20 & 50 & 50/900 (5.6) & $3-13$ \\
		10 & 2282 & 10 & 1323 $\pm$ 5 & 40 & 100 & 100/880 (11.4)  & $3-10$ \\
		11 & 2294 & 10 & 1299 $\pm$ 3 & 30 & 75 & 75/785 (9.6) & $3-10$ \\
		12 & 2298 & 9 & 1342 $\pm$ 5 & 30 & 75 & 75/770 (9.7) & $3-10$ \\
		13 & 2304 & 10 & 1316 $\pm$ 4 & 20 & 50 & 50/768 (6.8) & $3-13$\\
		14,15 & 2318 & 80 & 1320 $\pm$ 1 & 4 & 10 & 10/704 (1.4) & $3-13$ \\
		16 & 2340 & 11 & 1316 $\pm$ 5 & 30 & 75 & 75/731 (10.3) & $3-10$ \\
		17 & 2354 & 11 & 1327 $\pm$ 5 & 20 & 125 & 125/954 (13.1) & $3-20$\\
		18$^{e}$ & 2372 & 10 & - & - & - & - & $3-20$ \\ 	
		\hline
      \end{tabular}
      \begin{tablenotes}
       \item[e]Unable to constrain period
	  \end{tablenotes}
	 \end{threeparttable} 
\end{table*}
}
\begin{table}
	\begin{threeparttable}
	\caption{\small{Results of spectral fits using phenomenological model to source spectrum after subtraction of IGR spectrum is presented here. Spectral parameters are close to those in Table \ref{tab:pheno_fit} for corresponding observations. Similar variation in parameters is observed for all the observations.}}
	   \label{tab:app2}
	  \begin{tabular}{c@{\hspace{2pt}}c@{\hspace{2pt}}c@{\hspace{2pt}}c@{\hspace{2pt}}c@{\hspace{2pt}}c@{\hspace{2pt}}c} 
		\hline
		Obs & $N_{H}$ & $\Gamma$ & $E_{line}$ & $T_{in}$ & $Norm_{disk}$ & $\chi^{2}_{red}$ \\
		 \tiny{(ObsID)} & \tiny{($\times 10^{22}$ at/cm$^{2}$)} & & \tiny{(keV)} & \tiny{(keV)} & & \tiny{$\chi^2/dof$}\\
		\hline
		14 & $8.1^{+0.1}_{-0.1}$ & $2.9^{+0.1}_{-0.2}$ & - & $1.27^{+0.01}_{-0.01}$ & $273^{+13}_{-13}$ & 1.24 \\
		2318 & & & & & & (575/462)\\
		17 & $8.03^{+0.05}_{-0.09}$ & $2.14^{+0.01}_{-0.01}$ & $6.8^{l}$ & $1.27^{+0.01}_{-0.01}$ & $339^{+13}_{-13}$ & 1.04 \\
		2354 & & & & & & (452/431)\\
	   \hline
      \end{tabular}
      \begin{tablenotes}
       \item[l]Upper limit
	  \end{tablenotes}
	 \end{threeparttable}	  
\end{table}
To estimate the contamination due to IGR, we carried out pulsation analysis of the entire observations of \textit{AstroSat} of 2016 and 2018 outburst of 4U 1630-472. The \textit{LAXPC} lightcurves in $3-80$ keV energy band with a time bin of 1 sec were used to generate the $\chi^{2}$ plot with respect to the period. In order to find out the pulsation of the nearby source IGR, we used \textit{efsearch} tool of Xronos version 5.22. The default epoch of the observation was set for searching the period of 1310 sec \citep{Rodriguez2006}. The default phase bin per period was set to 10. Thus, it creates the bin time to 131 sec. The resolution for period search was set to 1.5 sec and the number of the search frequency was set to 32. These default parameters were used for searching the pulse period which gave the expected cycle of 68.97 with 690 newbins per interval for the ObsID 0698 (2016 outburst) and 142.03 expected cycle with 1421 newbins per interval for the ObsID 2318 (2018 outburst).

We find the pulsation varying from 1314 $\pm$ 2 s for ObsID 0698 to 1320 $\pm$ 1 s for ObsID 2318. Summary of the pulse period detected in other ObsIDs are summarized in Table \ref{tab:app1}.
\begin{figure}
	\includegraphics[width=\columnwidth]{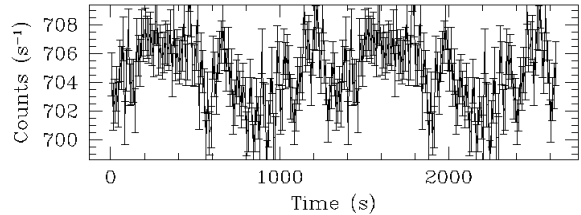}
    \caption{Pulse profile of ObsID 2318, showing pulse amplitude of 4 c/s.}
      \label{fig:A2}
\end{figure}
In order to find out the contribution of IGR J16320-4751 in each observation of 4U 1630-472, we calculated the pulse profile for the period around 1300 s. In Figure \ref{fig:A2}, we plot the flux variation of ObsID 2318 to show the pulse profile, which shows the amplitude of pulsation is 4 cps. The detected period is around 1314.5 sec, which is close to the earlier report in the literature. The pulse fraction of this source is reported in the range of $18-26$\% \citep{Rodriguez2006}. We adopt a nominal value of 20\% to get the total expected counts. Considering the offset for this pulse amplitude and assuming a pulse fraction of 20\%, this translates to about 10 c/s, while the total count rate is about 700 c/s (see Table \ref{tab:app1}). This gives the contamination of 1/70, which is very small and can possibly be neglected. Table \ref{tab:app1} gives the pulse period, pulse amplitude and typical contribution from the IGR source in each ObsID. The pulse amplitude is then used to obtain the relative contribution from the IGR source as a fraction of the total counts observed for that particular ObsID which is represented as the contamination percentage. Following similar approach, we estimate the contribution of IGR source for each observation which is presented in Table \ref{tab:app1}.

The source IGR has a hard spectrum as compared to 4U 1630-472 and as a result even with this low counts, it can contribute to some extent at high energies. To account for this we scale the simulated spectrum of IGR to give the estimated count rate using \textit{LAXPC} response for an offset of 0.5$^{\circ}$. This spectrum is compared with the observed spectrum and we restrict our analysis to the range where the contribution of contaminating source is less than 5\%. As a further check we do two fits, one using the observed spectrum and another after subtracting the contribution of the contaminating source to check if the effect is significant. This two pronged approach helps us to determine the energy range that can be considered absolutely contamination free in the worst case scenario. The parameters obtained for two sample cases in ObsID 2318 and 2354 are given in Table \ref{tab:app2}. Comparing with the parameters in Table \ref{tab:pheno_fit}, it can be seen that the spectral fits with and without the IGR spectrum subtracted are comparable. Based on these studies, the energy band considered for each ObsID is given in the last column of Table \ref{tab:app1}.


\bsp	
\label{lastpage}
\end{document}